\documentclass[lettersize,journal]{IEEEtran}
\usepackage{amsmath,amsfonts}
\usepackage{amsthm}
\usepackage{amssymb}
\usepackage{mathtools}

\usepackage{algorithmic}
\usepackage{algorithm}
\usepackage{subcaption}

\usepackage{array}

\usepackage{textcomp}
\usepackage{stfloats}
\usepackage{url}
\usepackage{verbatim}
\usepackage{graphicx}
\usepackage{cite}
\usepackage{cuted}

\DeclareMathSizes{10}{9}{6}{5} %表示正文为 10 pt 时，公式主体、上下标、二级上下标分别使用 9、6、5 pt

\allowdisplaybreaks[2]

\usepackage{bm}
\graphicspath{{figures/}}

\usepackage{color}

\newlength{\alguntilwidth}
\begin{document}

\title{{\fontsize{24pt}{22pt}\selectfont 
Active-Element Configuration for Energy-Efficient Secure Near-Field ISAC With Hybrid RISs}}
  %Energy-Efficient Secure Transmission for Hybrid-RIS-Assisted Near-Field ISAC      
\author{Ling Zhuang,\IEEEmembership{~IEEE Student Member,} Ximing Xie,\IEEEmembership{~IEEE Member,} Fang Fang,\IEEEmembership{~IEEE Senior Member,} \\ and Zhizhong Zhang, \IEEEmembership{~IEEE Senior Member}
        % <-this % stops a space
 %\thanks{This work was supported in part by the Major Science and Technology Project of Jiangsu Province under Grant BG2024002; in part by the Natural Science Foundation of China under Grant 62401271.  \textit{(Corresponding authors: Fang Fang; Zhizhong Zhang.)}}
 \vspace{-2mm}
 
\thanks{Ling Zhuang and Zhizhong Zhang are with the School of Electronic and Information Engineering, Nanjing University of Information Science and Technology, Nanjing, 210044, China (e-mail: 20211118008@nuist.edu.cn; zhangzz@nuist.edu.cn). Ximing Xie is with the Department of Electrical and Computer Engineering, Western University, London, ON, N6A 3K7, Canada (e-mail: xxie269@uwo.ca). Fang Fang is with the Department of Electrical and Computer Engineering and the Department of Computer Science, Western University, London, ON, N6A 3K7, Canada (e-mail: fang.fang@uwo.ca).}
% <-this % stops a space
}



\maketitle

\begin{abstract}
Hybrid active-passive reconfigurable intelligent surfaces (HRISs) are promising for secure near-field integrated sensing and communication (ISAC) through selective signal amplification and spherical-wave beam focusing. However, fixed active-element layouts limit flexibility in exploiting element-dependent near-field channels while balancing secrecy performance, sensing accuracy, and circuit power consumption. This paper develops a joint active-element configuration and transmission design for a large-aperture HRIS-assisted system, where sensing targets act as potential eavesdroppers. The design adapts both the number and spatial distribution of active elements by exploiting near-field phase and attenuation variations. We establish a bidirectional signal and power model with full-aperture reflection and active-subset amplification for downlink signals and uplink echoes. Based on the derived three-dimensional localization Cram\'er-Rao bound, we formulate a mixed-integer non-convex secrecy energy efficiency (SEE) maximization problem under user-rate, localization, and power constraints. To solve this problem, we propose an alternating optimization framework combining successive convex approximation, a penalty convex-concave procedure, and full-aperture reweighted group-sparse selection. The selection method scores and ranks elements to generate binary candidate configurations without exhaustive search. After refining continuous HRIS variables for each candidate, we select the feasible configuration with the highest genuine SEE. Simulation results demonstrate higher SEE with lower base station transmit power than fixed-layout HRIS schemes and a fully passive RIS baseline.
\end{abstract}
\vspace{-1mm}
\begin{IEEEkeywords}
Reconfigurable intelligent surfaces, physical layer security, integrated sensing and communication, near field.
\end{IEEEkeywords}
%%%%%%%%%%%%%%%%%%%%%%%%%%%%%%%%%%%%%%%%%%%%
%%%%%%%%%%%%%%%%%%%%%%%%%%%%%%%%%%%%%%%%%%%%
%%%%%%%%%%%%%%%%%%%%%%%%%%%%%%%%%%%%%%%%%%%%
%%%%%%%%%%%%%%%%%%%%%%%%%%%%%%%%%%%%%%%%%%%%
\vspace{-4mm}
\section{Introduction}
\IEEEPARstart{S}{ixth}-generation wireless networks are expected to support high-rate communication and precise environmental sensing. To meet these requirements, high-frequency bands provide abundant spectral resources for integrated sensing and communication (ISAC)~\cite{r1}, but suffer from severe propagation loss. To compensate for this loss, large-aperture arrays are typically employed. These large apertures, together with short wavelengths, extend the near-field region, where spherical-wave propagation enables beam focusing in both range and angle for communication and target illumination~\cite{r2}. However, sensing targets may also act as eavesdroppers (Eves) and intercept confidential information intended for legitimate users. Therefore, secure ISAC requires ensuring sufficient target illumination for sensing while limiting information leakage toward the targets. Physical-layer security has emerged as an attractive solution for secure ISAC, which can exploit spatial channel differences to enhance system secrecy performance~\cite{r3}.

However, in practical secure ISAC systems, when the direct links from the base station (BS) to users and targets are blocked, the remaining non-line-of-sight (NLoS) paths may be insufficient for reliable communication and sensing. To overcome this limitation, reconfigurable intelligent surfaces (RISs) provide controllable reflection paths to bypass blockage and extend coverage~\cite{r5}. In RIS-assisted secure ISAC, the surface reflects downlink (DL) signals toward users and targets and returns uplink (UL) target echoes to the BS~\cite{r6}. Compared with passive RISs, active RISs can mitigate cascaded attenuation by amplifying both DL signals and UL echoes, but introduce additional noise and power consumption~\cite{r7}. For large-aperture RISs, operating all elements in active mode can incur substantial circuit power consumption. Hybrid active-passive RISs (HRISs) reduce this cost by activating only a subset of elements~\cite{r8}. However, some existing studies on HRIS-assisted secure ISAC optimize the BS transmit beamformers and HRIS phase shifts, while keeping the active-element configuration fixed~\cite{r9,r10}. Such a fixed configuration limits the flexibility to balance secrecy performance, sensing accuracy, and circuit power consumption. This limitation is particularly important in the near field, where the propagation distances from different HRIS elements to each recipient vary across the aperture, resulting in element-dependent channel phase and attenuation. Consequently, active elements at different positions make different contributions to communication and sensing; thus, a fixed active-element spatial distribution may limit the ability to exploit these element-wise channel variations.

These observations motivate the joint optimization of active-element configuration and transmission for HRIS-assisted secure near-field ISAC. By adapting the number and spatial distribution of active elements to the element-dependent near-field channels, the HRIS can allocate amplification resources more effectively, thereby improving the balance among secrecy performance, sensing accuracy, and total power consumption.

\vspace{-3mm}
\subsection{Related Works}
Active and passive RISs have been investigated for secure transmission and ISAC across various propagation scenarios~\cite{r11,r13,r14}. Hybrid surface architectures have also been explored to combine active amplification with passive reflection. The authors of~\cite{r10} investigated a multi-HRIS-assisted system for joint DL communication and target sensing. In their model, an extended target acted as a potential Eve. Considering the uncertainty in the target location, they jointly optimized the transceiver beamforming and HRIS coefficients to maximize the worst-case sensing SINR under communication, security, and power constraints. In~\cite{r15}, the authors adopted a sub-connected active RIS architecture in a millimeter-wave ISAC system, where multiple elements shared a single amplifier. Hybrid precoding at the BS and beamforming at the RIS were jointly optimized to maximize secrecy energy efficiency while satisfying communication and sensing requirements. Furthermore, advanced hybrid surface architectures and near-field propagation have been exploited to enhance localization and security. In~\cite{r16}, a semi-passive RIS employed passive elements for signal reflection and receive sensors for echo reception. The authors optimized the BS and RIS beamformers to minimize the localization Cram\'er--Rao bound (CRB) under user-rate constraints. Regarding secure near-field ISAC,~\cite{r17} exploited spherical-wave beam focusing to distinguish a legitimate user from an eavesdropping target aligned in the same angular direction, thereby achieving a positive secrecy rate through range-domain resolution. These studies mainly optimize continuous transmission and reception variables under fixed array or surface configurations.

To improve the configuration flexibility of HRISs, several studies have optimized the number and spatial distribution of active elements. In~\cite{r18}, the numbers of active and passive elements were optimized to maximize the worst-user ergodic capacity in multiuser communications under Rician fading, subject to deployment-cost and amplification-power constraints. For a given number of active elements, the authors of~\cite{r19} investigated their spatial distribution in an HRIS-assisted multiple-input single-output (MISO) communication system. Their results demonstrated that optimizing the spatial distribution of active elements according to channel conditions improved the received signal-to-noise ratio (SNR). Adaptive active-element configurations have also been investigated for sensing and secure transmission. In~\cite{r20}, an HRIS assisted target illumination and echo reception in a multiuser ISAC system. The element-wise active/passive mode selection, transmit beamforming, RIS coefficients, and the radar receive filter were jointly optimized to maximize the radar output SNR under user SINR and power constraints. For HRIS-assisted secure communications, the authors of~\cite{r21} considered fixed and dynamic HRIS architectures in a MISO system with a multi-antenna Eve. In the dynamic architecture, they jointly optimized the BS transmit covariance, HRIS reflection coefficients, and element-wise active/passive mode selection to maximize the secrecy rate. This mode selection allows both the number and spatial distribution of active elements to adapt to channel conditions. 
\vspace{-3mm}
\subsection{Motivation and Contributions}
In secure near-field ISAC systems, the active-element configuration has a coupled impact on legitimate reception, information leakage, localization accuracy, the amplification noise, and power consumption in both DL and UL transmission. Unlike \cite{r20}, which considers HRIS mode selection for ISAC without secure near-field localization, and \cite{r21}, which studies dynamic HRIS selection for secure communications without sensing localization or UL echo reception, our setting further involves element-dependent phase and attenuation variations across the large HRIS aperture. Consequently, the number and spatial distribution of active elements jointly affect secrecy transmission, bidirectional sensing, and power consumption. Motivated by this coupling, we jointly optimize the active-element configuration and transmission variables to maximize the SEE in a large-aperture HRIS-assisted secure near-field ISAC system.
The main contributions of this paper are summarized as follows:
\vspace{-1mm}
\begin{itemize}
\item
We investigate the joint active-element configuration and transmission design for a large-aperture HRIS-assisted secure near-field ISAC system, where sensing targets act as potential Eves. Subject to user-rate, CRB-based localization accuracy, and power constraints, we formulate a SEE maximization problem to balance secrecy performance and total power consumption. Specifically, the number and spatial distribution of active elements are adapted by exploiting the element-wise phase and attenuation variations in near-field channels. This adaptation also accounts for the amplification gains, amplifier noise, and power consumption in both DL and UL branches. 

\item
We establish a bidirectional near-field signal model for DL transmission and UL echo reception, accounting for full-aperture reflection and active-subset amplification. The unknown target reflection coefficients are treated as nuisance parameters to derive the three-dimensional (3D) localization CRB matrix for multi-target position estimation. The CRB constraint is reformulated as a
semidefinite matrix inequality via the Schur complement, allowing the localization requirement to be incorporated into the joint configuration and transmission design.

\item
To address the mixed-integer non-convex problem, we develop an alternating optimization (AO) framework based on successive convex approximation (SCA). The active and passive phase shifts are optimized directly in vector form using a penalty convex-concave procedure (PCCP). Furthermore, we design a full-aperture reweighted group-sparse selection method to determine the active-element configuration without exhaustive search. This method scores and ranks the HRIS elements by solving a convex surrogate problem to generate candidate configurations and recover them in binary form. The continuous HRIS variables are then refined for each candidate. The resulting configuration is accepted only if it satisfies the original constraints and improves the genuine SEE.

\item
Numerical results demonstrate that the proposed design achieves higher SEE with lower BS transmit power than fixed-layout HRIS schemes with the same number of active elements and a fully passive RIS baseline. The results also reveal a tradeoff between passive aperture gain and active amplification. These findings highlight the benefits of adapting the number and spatial distribution of active elements to balance secure transmission, localization accuracy, and power consumption.

\end{itemize}
%\subsection{Organization}
%The remainder of this paper is organized as follows. Section~II presents the system model and formulates the SEE maximization problem. Section~III develops the proposed AO framework and analyzes its computational complexity and convergence. Section~IV presents the simulation results. Finally, Section~V concludes the paper.
\vspace{-2mm}
\textit{Notation:} $\mathbb{S}_+^n$ denotes the set of $n\times n$ real symmetric positive semidefinite matrices. $\mathbf{I}_n$ and $\mathbf{1}_M$ represent the $n\times n$ identity matrix and $M$-dimensional all-ones vector, respectively. The operators $\operatorname{vec}(\cdot)$ and $\odot$ denote column-wise vectorization and the Hadamard product, respectively. Finally, $[x]^+\triangleq\max\{0,x\}$.

\begin{figure}[!t]
\centering
\includegraphics[width=2.2in, trim=0 0 0 0, clip]{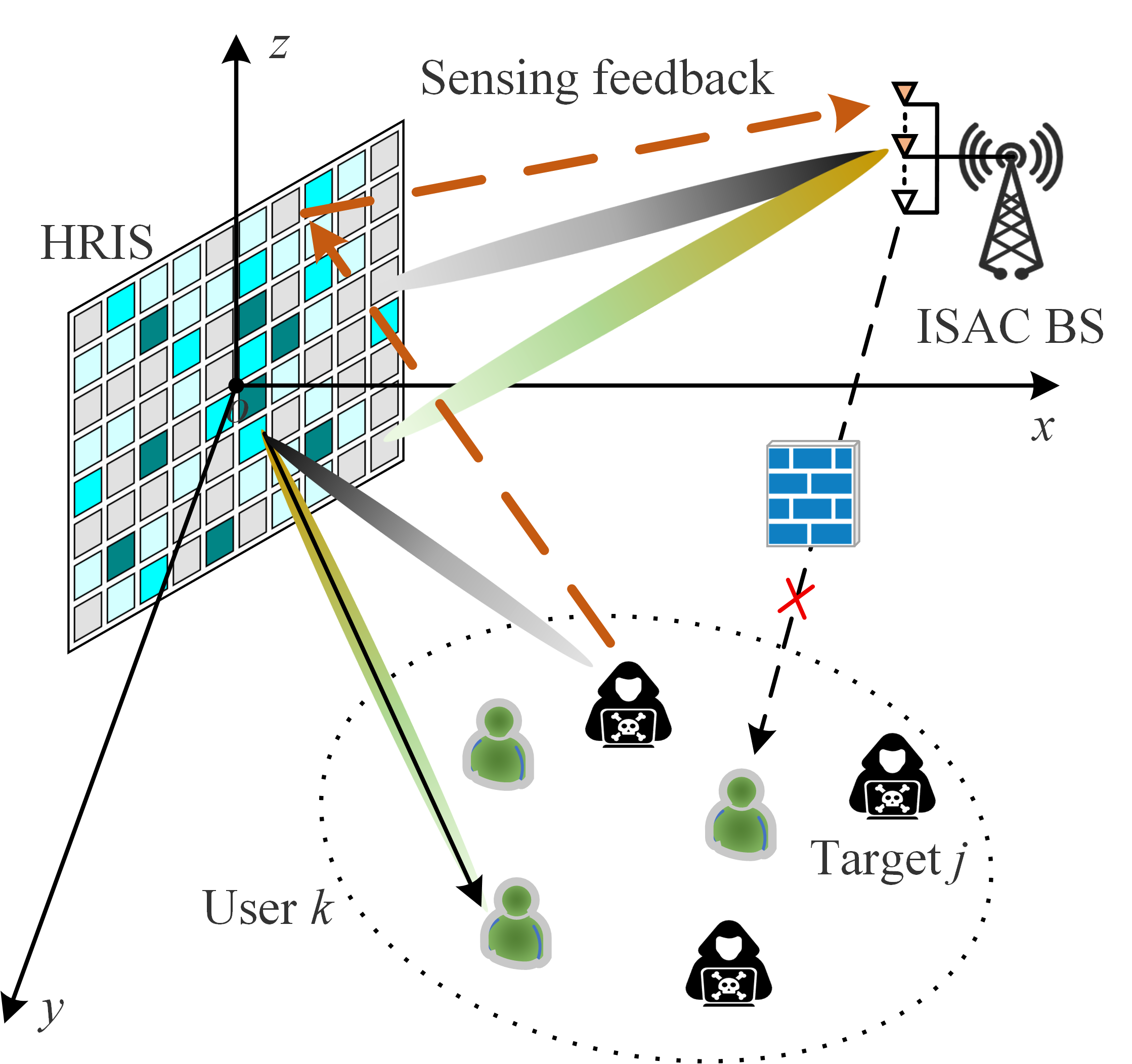}
\caption{The HRIS-assisted secure ISAC in the near field.}
\label{fig1}
\vspace{-6mm}
\end{figure}
%%%%%%%%%%%%%%%%%%%%%%%%%%%%%%%%%%%%%%%%%%%%
%%%%%%%%%%%%%%%%%%%%%%%%%%%%%%%%%%%%%%%%%%%%
%%%%%%%%%%%%%%%%%%%%%%%%%%%%%%%%%%%%%%%%%%%%
%%%%%%%%%%%%%%%%%%%%%%%%%%%%%%%%%%%%%%%%%%%%\vspace{-2mm}
\section{System Model and Problem Formulation}

As shown in Fig.~\ref{fig1}, we consider an HRIS-assisted secure ISAC system, where the BS simultaneously serves $K$ single-antenna legitimate users and senses $J$ targets that also act as potential Eves. The BS, users, and Eves are assumed to lie in the near field of the HRIS, with their distances from the HRIS center falling within the Fraunhofer distance $R_{\mathrm{F}}=\text{2}D^{\text{2}}/\lambda_c$~\cite{r22}, where $D$ denotes the maximum distance between HRIS elements and $\lambda_c$ is the carrier wavelength. The direct BS-to-user and BS-to-target links are neglected due to severe blockage. We use a 3D Cartesian coordinate system with the HRIS centered at the origin.
The BS employs an $N$-element uniform linear array parallel to the $z$-axis and centered at $\boldsymbol{r}_{\mathrm{b}}=\left[x_{\mathrm{b}},y_{\mathrm{b}},z_{\mathrm{b}}\right]^{T}$. Its $n$-th antenna is located at $\boldsymbol{r}_{\mathrm{b},n}=\left[x_{\mathrm{b}},y_{\mathrm{b}},z_{\mathrm{b}}-\Delta_n^N d\right]^{T}$, where $\Delta_i^I=i-\text{1}-\left(I-\text{1}\right)/\text{2}$. The HRIS comprises $M=M_y\times M_z$ elements arranged in a uniform planar array, with $M_y$ and $M_z$ elements along the $y$- and $z$-axes, respectively. The position of the $m$-th element is $\boldsymbol{r}_{\mathrm{ris},m}=\left[\text{0},\Delta_{m_y}^{M_y}d,\Delta_{m_z}^{M_z}d\right]^{T}$, where $m=\left(m_z-\text{1}\right)M_y+m_y$, $m\in\mathcal{M}=\left\{\text{1},\ldots,M\right\}$, $m_y\in\left\{\text{1},\ldots,M_y\right\}$, and $m_z\in\left\{\text{1},\ldots,M_z\right\}$. Both arrays adopt an inter-element spacing of $d=\lambda_c/\text{2}$. User $k$ and Eve $j$ are located at $\boldsymbol{r}_{\mathrm{u},k}=\left[x_{\mathrm{u},k},y_{\mathrm{u},k},z_{\mathrm{u},k}\right]^{T}$ and $\boldsymbol{r}_{\mathrm{e},j}=\left[x_{\mathrm{e},j},y_{\mathrm{e},j},z_{\mathrm{e},j}\right]^{T}$, respectively, for $k\in\mathcal{K}=\left\{\text{1},\ldots,K\right\}$ and $j\in\mathcal{J}=\left\{\text{1},\ldots,J\right\}$.

The HRIS supports both DL transmission and UL echo reception. Each element can switch between active and passive modes. Both modes support signal reflection, while the selected active elements provide controllable amplification and independent phase adjustment for the DL and UL branches. To characterize the number and spatial distribution of active elements, we define the selection vector $\boldsymbol{e}=\left[e_{\text{1}},\ldots,e_M\right]^{T}\in\left\{\text{0},\text{1}\right\}^{M\times\text{1}}$, where the indicator $e_m$ is given by
\begin{equation}
e_m=
\begin{cases}
\text{1}, & \text{if element $m$ operates in active mode},\\
\text{0}, & \text{if element $m$ operates in passive mode}.
\end{cases}
\end{equation}
In passive mode, the amplification circuit is bypassed. In active mode, each element uses a shared reflection-type amplifier and an RF splitter, with separate phase control for the two branches \cite{r24}. The sufficient isolation between the branches ensures negligible internal cross-coupling. Let $\boldsymbol{\alpha}=\left[\alpha_{\text{1}},\ldots,\alpha_M\right]^{T}\in\mathbb{R}^{M\times\text{1}}$ denote the amplitude amplification vector, with $\text{1}\leq\alpha_m\leq\alpha_{\max}$. The power-splitting vector is $\boldsymbol{\rho}=\left[\rho_{\text{1}},\ldots,\rho_M\right]^{T}\in\left[\text{0},\text{1}\right]^{M\times\text{1}}$, where $\rho_m$ represents the fraction of power gain allocated to the DL branch, while the remaining fraction is allocated to the UL branch. Therefore, the effective DL and UL amplitude gains are $\alpha_m\sqrt{\rho_m}$ and $\alpha_m\sqrt{\text{1}-\rho_m}$, respectively. For a selected active element, setting $\rho_m=\text{0}$ or $\rho_m=\text{1}$ deactivates its active DL or UL branch, respectively.

Let $\boldsymbol{\phi}_{\mathrm{a}}^{\mathrm{c}}=\left[e^{\mathrm{j}\phi_{\mathrm{a},\text{1}}^{\mathrm{c}}},\ldots,e^{\mathrm{j}\phi_{\mathrm{a},M}^{\mathrm{c}}}\right]^{T}$ and $\boldsymbol{\phi}_{\mathrm{a}}^{\mathrm{s}}=\left[e^{\mathrm{j}\phi_{\mathrm{a},\text{1}}^{\mathrm{s}}},\ldots,e^{\mathrm{j}\phi_{\mathrm{a},M}^{\mathrm{s}}}\right]^{T}$ denote the phase-shift vectors for the active DL and UL branches, respectively, where  $\phi_{\mathrm{a},m}^{\mathrm{c}},\phi_{\mathrm{a},m}^{\mathrm{s}}\in\left[\text{0},\text{2}\pi\right)$. The effective active response vectors incorporate the active-element selection indicators, branch amplitude gains, and phase shifts, which are formulated as
\begin{equation}
\bar{\boldsymbol{\phi}}_{\mathrm{a}}^{\mathrm{c}}
=\boldsymbol{e}\odot\sqrt{\boldsymbol{\rho}}
\odot\boldsymbol{\alpha}\odot\boldsymbol{\phi}_{\mathrm{a}}^{\mathrm{c}},
\label{DL active response}
\end{equation}
\begin{equation}
\bar{\boldsymbol{\phi}}_{\mathrm{a}}^{\mathrm{s}}
=\boldsymbol{e}\odot\sqrt{\mathbf{\text{1}}_M-\boldsymbol{\rho}}
\odot\boldsymbol{\alpha}\odot\boldsymbol{\phi}_{\mathrm{a}}^{\mathrm{s}}.
\label{UL active response}
\end{equation}
The active-element parameters are defined over all $M$ HRIS positions, and $\boldsymbol{e}$ ensures that only the selected elements contribute to the active responses. The corresponding response matrices are $\boldsymbol{\Phi}_{\mathrm{a}}^{\mathrm{c}}=\mathrm{diag}\left(\bar{\boldsymbol{\phi}}_{\mathrm{a}}^{\mathrm{c}}\right)$ and $\boldsymbol{\Phi}_{\mathrm{a}}^{\mathrm{s}}=\mathrm{diag}\left(\bar{\boldsymbol{\phi}}_{\mathrm{a}}^{\mathrm{s}}\right)$. The unselected elements operate in passive mode and use the same phase-shift vector for DL and UL reflection, defined as $\boldsymbol{\phi}_{\mathrm{p}}=\left[e^{\mathrm{j}\phi_{\mathrm{p},\text{1}}},\ldots,e^{\mathrm{j}\phi_{\mathrm{p},M}}\right]^{T}$ with $\phi_{\mathrm{p},m}\in\left[\text{0},\text{2}\pi\right)$. The full-aperture DL and UL response vectors are formulated as
\begin{equation}
\boldsymbol{\theta}_{\mathrm{c}}
=\bar{\boldsymbol{\phi}}_{\mathrm{a}}^{\mathrm{c}}
+\left(\mathbf{\text{1}}_M-\boldsymbol{e}\right)
\odot\boldsymbol{\phi}_{\mathrm{p}},
\end{equation}
\begin{equation}
\boldsymbol{\theta}_{\mathrm{s}}
=\bar{\boldsymbol{\phi}}_{\mathrm{a}}^{\mathrm{s}}
+\left(\mathbf{\text{1}}_M-\boldsymbol{e}\right)
\odot\boldsymbol{\phi}_{\mathrm{p}}.
\end{equation}
Then, the complete HRIS response matrices are $\boldsymbol{\Theta}_{\mathrm{c}}=\mathrm{diag}\left(\boldsymbol{\theta}_{\mathrm{c}}\right)$ and $\boldsymbol{\Theta}_{\mathrm{s}}=\mathrm{diag}\left(\boldsymbol{\theta}_{\mathrm{s}}\right)$ for the DL and UL branches, respectively.

The full-duplex BS simultaneously transmits ISAC signals and receives target echoes over the same time-frequency resources. The communication symbol vector is $\boldsymbol{s}_{\mathrm{c}}=\left[s_{\mathrm{c},\text{1}},\ldots,s_{\mathrm{c},K}\right]^{T}\in\mathbb{C}^{K\times\text{1}}$  with $\mathbb{E}\left(\boldsymbol{s}_{\mathrm{c}}\boldsymbol{s}_{\mathrm{c}}^{H}\right)=\mathbf{I}_K$. The sensing symbol vector is $\boldsymbol{s}_{\mathrm{s}}=\left[s_{\mathrm{s},\text{1}},\ldots,s_{\mathrm{s},N}\right]^{T}\in\mathbb{C}^{N\times\text{1}}$ with $\mathbb{E}\left(\boldsymbol{s}_{\mathrm{s}}\boldsymbol{s}_{\mathrm{s}}^{H}\right)=\mathbf{I}_N$. Both vectors are zero-mean and mutually independent. Let $\mathbf{W}_{\mathrm{c}}=\left[\boldsymbol{w}_{\mathrm{c},\text{1}},\ldots,\boldsymbol{w}_{\mathrm{c},K}\right]\in\mathbb{C}^{N\times K}$ and $\mathbf{W}_{\mathrm{s}}=\left[\boldsymbol{w}_{\mathrm{s},\text{1}},\ldots,\boldsymbol{w}_{\mathrm{s},N}\right]\in\mathbb{C}^{N\times N}$ denote the communication and sensing beamforming matrices, respectively. The transmitted signal is $\mathbf{x}=\mathbf{W}_{\mathrm{c}}\boldsymbol{s}_{\mathrm{c}}+\mathbf{W}_{\mathrm{s}}\boldsymbol{s}_{\mathrm{s}}$. 

%%%%%%%%%%%%%%%%%%%%%%%%%%%%%%%%
\vspace{-2mm}
\subsection{Near-Field Secure Communication Model}

To characterize the signal propagation in high-frequency bands, we model the HRIS-assisted links as near-field Rician channels~\cite{r23}. The DL BS-to-HRIS channel $\boldsymbol{G}\in\mathbb{C}^{M\times N}$ is expressed as
\begin{equation}
\boldsymbol{G}
=\sqrt{\beta_{\mathrm{b}}}\boldsymbol{\Upsilon}_{\mathrm{b}}
\odot\left(
\sqrt{\frac{\kappa_{\mathrm{b}}}{\text{1}+\kappa_{\mathrm{b}}}}
\boldsymbol{G}^{\mathrm{LoS}}
+\sqrt{\frac{\text{1}}{\text{1}+\kappa_{\mathrm{b}}}}
\boldsymbol{G}^{\mathrm{NLoS}}
\right),
\end{equation}
where $\beta_{\mathrm{b}}>\text{0}$ and $\kappa_{\mathrm{b}}>\text{0}$ denote the distance-independent gain and Rician factor, respectively. The attenuation matrix $\boldsymbol{\Upsilon}_{\mathrm{b}}\in\mathbb{R}^{M\times N}$ accounts for free-space spreading and molecular absorption losses. Its $\left(m,n\right)$-th entry is
$\left[\boldsymbol{\Upsilon}_{\mathrm{b}}\right]_{m,n}
=\frac{\lambda_c}{\text{4}\pi d_{\mathrm{b},m,n}}
e^{-\frac{\text{1}}{\text{2}}\mu_{\mathrm{abs}}d_{\mathrm{b},m,n}}$,
where
$d_{\mathrm{b},m,n}
=\left\Vert
\boldsymbol{r}_{\mathrm{b},n}
-\boldsymbol{r}_{\mathrm{ris},m}
\right\Vert$,
and $\mu_{\mathrm{abs}}$ denotes the molecular absorption coefficient at the carrier frequency and under the considered atmospheric conditions. The LoS component has entries
$\left[\boldsymbol{G}^{\mathrm{LoS}}\right]_{m,n}
=e^{-\mathrm{j}\frac{\text{2}\pi}{\lambda_c}d_{\mathrm{b},m,n}}$,
while the NLoS entries satisfy
$\left[\boldsymbol{G}^{\mathrm{NLoS}}\right]_{m,n}
\sim\mathcal{CN}\left(\text{0},\text{1}\right)$.

Similarly, the HRIS-to-user channel
$\boldsymbol{h}_{\mathrm{u},k}\in\mathbb{C}^{M\times\text{1}}$
and the HRIS-to-Eve channel
$\boldsymbol{h}_{\mathrm{e},j}\in\mathbb{C}^{M\times\text{1}}$
are modeled as
\begin{equation}
\boldsymbol{h}_{p,q}=\sqrt{\beta_{p,q}}\boldsymbol{\Upsilon}_{p,q}
\odot\left(
\sqrt{\frac{\kappa_{p,q}}{\text{1}\!+\!\kappa_{p,q}}}
\boldsymbol{h}_{p,q}^{\mathrm{LoS}}
\!+\!\sqrt{\frac{\text{1}}{\text{1}\!+\!\kappa_{p,q}}}
\boldsymbol{h}_{p,q}^{\mathrm{NLoS}}
\right),
\end{equation}
where $p\in\left\{\mathrm{u},\mathrm{e}\right\}$, with $q=k\in\mathcal{K}$ for $p=\mathrm{u}$ and $q=j\in\mathcal{J}$ for $p=\mathrm{e}$. The parameters $\beta_{p,q}>\text{0}$ and $\kappa_{p,q}>\text{0}$ denote the distance-independent gain and Rician factor, respectively. The attenuation vector
$\boldsymbol{\Upsilon}_{p,q}\in\mathbb{R}^{M\times\text{1}}$
has entries
$\left[\boldsymbol{\Upsilon}_{p,q}\right]_{m}
=\frac{\lambda_c}{\text{4}\pi d_{p,m,q}}
e^{-\frac{\text{1}}{\text{2}}\mu_{\mathrm{abs}}d_{p,m,q}}$,
where
$d_{p,m,q}
=\left\Vert
\boldsymbol{r}_{\mathrm{ris},m}-\boldsymbol{r}_{p,q}
\right\Vert$.
The corresponding LoS entries are
$\left[\boldsymbol{h}_{p,q}^{\mathrm{LoS}}\right]_{m}
=e^{-\mathrm{j}\frac{\text{2}\pi}{\lambda_c}d_{p,m,q}}$,
and the NLoS entries satisfy
$\left[\boldsymbol{h}_{p,q}^{\mathrm{NLoS}}\right]_{m}
\sim\mathcal{CN}\left(\text{0},\text{1}\right)$.
Hence, the DL signal received at user $k$ or Eve $j$ is given by
\begin{equation}
y_{p,q}=\boldsymbol{h}_{p,q}^{T}
\boldsymbol{\Theta}_{\mathrm{c}}\boldsymbol{G}\mathbf{x}+\boldsymbol{h}_{p,q}^{T}
\boldsymbol{\Phi}_{\mathrm{a}}^{\mathrm{c}}
\boldsymbol{n}_{\mathrm{a}}^{\mathrm{c}}
+n_{p,q},
\end{equation}
where
$\boldsymbol{n}_{\mathrm{a}}^{\mathrm{c}}
\sim\mathcal{CN}\left(
\mathbf{\text{0}},
\delta_{\mathrm{a}}^{\text{2}}\mathbf{I}_{M}
\right)$
denotes the DL amplifier noise, and
$n_{p,q}\sim\mathcal{CN}\left(
\text{0},\delta_{p,q}^{\text{2}}
\right)$
denotes the receiver noise. These noises are assumed mutually independent and independent of the transmitted symbols. Since $\boldsymbol{\Phi}_{\mathrm{a}}^{\mathrm{c}}$ incorporates the active-element selection vector $\boldsymbol{e}$, only the selected active elements contribute to the amplifier noise term.

To decode stream $s_{\mathrm{c},k}$, user $k$ and Eve $j$ treat the remaining communication streams and all sensing streams as interference. In particular, the sensing symbols are assumed to be unavailable to the Eves for coherent cancellation. Therefore, the SINR at receiver $\left(p,q\right)$ for decoding stream $k$ is given by
\begin{equation}
\gamma_{p,q,k}
=\frac{
\left|
\boldsymbol{h}_{p,q}^{T}
\boldsymbol{\Theta}_{\mathrm{c}}
\boldsymbol{G}\boldsymbol{w}_{\mathrm{c},k}
\right|^{\text{2}}
}{
c_{p,q,k}
+\delta_{\mathrm{a}}^{\text{2}}
\left\Vert
\boldsymbol{h}_{p,q}^{T}
\boldsymbol{\Phi}_{\mathrm{a}}^{\mathrm{c}}
\right\Vert^{\text{2}}
+\delta_{p,q}^{\text{2}}
},
\label{eq:SINR}
\end{equation}
where the total interference power is
\begin{equation}
c_{p,q,k}
=\sum_{\substack{i\neq k}}^{K}
\left|
\boldsymbol{h}_{p,q}^{T}
\boldsymbol{\Theta}_{\mathrm{c}}
\boldsymbol{G}\boldsymbol{w}_{\mathrm{c},i}
\right|^{\text{2}}+\sum_{n=\text{1}}^{N}
\left|
\boldsymbol{h}_{p,q}^{T}
\boldsymbol{\Theta}_{\mathrm{c}}
\boldsymbol{G}\boldsymbol{w}_{\mathrm{s},n}
\right|^{\text{2}}.
\end{equation}
For user $k$, we use the shorthand notations
$\gamma_{\mathrm{u},k}\equiv\gamma_{\mathrm{u},k,k}$
and
$c_{\mathrm{u},k}\equiv c_{\mathrm{u},k,k}$.
For Eve $j$, the corresponding SINR and interference power for decoding stream $k$ are denoted by
$\gamma_{\mathrm{e},j,k}$ and $c_{\mathrm{e},j,k}$, respectively.
Then, the achievable rates of user $k$ and Eve $j$ for decoding stream $k$ are 
$R_{\mathrm{u},k}
=\log_{\text{2}}\left(
\text{1}+\gamma_{\mathrm{u},k}
\right)$
and
$R_{\mathrm{e},j,k}
=\log_{\text{2}}\left(
\text{1}+\gamma_{\mathrm{e},j,k}
\right)$,
respectively.

\vspace{-3mm}
\subsection{Near-Field Sensing Model}

At the same carrier frequency, we assume reciprocity of the BS-HRIS propagation channel and model the UL HRIS-to-BS channel as  
$\bar{\boldsymbol{G}}=\boldsymbol{G}^{T}$.
Let $\varrho_{\mathrm{e},j}\in\mathbb{C}$ denote the unknown normalized complex reflection coefficient of target $j$. The sensing signal $\boldsymbol{y}_{\mathrm{s}}\!\in\!\mathbb{C}^{N\times\text{1}}$ received at the BS is expressed as
\begin{equation}
\boldsymbol{y}_{\mathrm{s}}
=\sum_{j=\text{1}}^{J}
\varrho_{\mathrm{e},j}
\bar{\boldsymbol{G}}\boldsymbol{\Theta}_{\mathrm{s}}
\boldsymbol{h}_{\mathrm{e},j}
\boldsymbol{h}_{\mathrm{e},j}^{T}\left(
\boldsymbol{\Theta}_{\mathrm{c}}\boldsymbol{G}\mathbf{x}
+\boldsymbol{\Phi}_{\mathrm{a}}^{\mathrm{c}}
\boldsymbol{n}_{\mathrm{a}}^{\mathrm{c}}
\right)+\bar{\boldsymbol{G}}
\boldsymbol{\Phi}_{\mathrm{a}}^{\mathrm{s}}
\boldsymbol{n}_{\mathrm{a}}^{\mathrm{s}}
+\boldsymbol{n}_{\mathrm{BR}},
\end{equation}
where
$\boldsymbol{n}_{\mathrm{a}}^{\mathrm{s}}
\sim\mathcal{CN}\left(
\mathbf{\text{0}},
\delta_{\mathrm{a}}^{\text{2}}\mathbf{I}_{M}
\right)$
denotes the amplifier noise of the UL active branch.
The first term represents the target echoes of the transmitted waveform and DL amplifier noise, which are reflected by the full HRIS aperture back to the BS.
The second term represents the UL amplifier noise contributed by the selected active elements. The remaining disturbance at the BS is modeled as
$\boldsymbol{n}_{\mathrm{BR}}
\sim\mathcal{CN}\left(
\mathbf{\text{0}},\mathbf{R}_{\mathrm{BR}}
\right)$,
where
$\mathbf{R}_{\mathrm{BR}}\succ\mathbf{\text{0}}$
includes receiver thermal noise, residual full-duplex self-interference, environmental clutter, and unmodeled scattering.

%%%%%%%%%%%%%%%%%%%%%%%%%%%%%%%%
\vspace{-4mm}

\subsection{CRB for Near-Field 3D Localization}

To quantify near-field 3D localization accuracy, we derive the CRB from the received sensing signal under the noise approximation specified below. The CRB provides a lower bound on the error covariance of any unbiased estimator under the adopted observation model~\cite{r25}. We consider a coherent processing interval (CPI) of $L$ snapshots, during which the HRIS configuration and channel parameters are assumed to be time-invariant. The target positions and complex reflection coefficients are treated as deterministic unknowns and collected in a real parameter vector $\boldsymbol{\zeta}=\left[\boldsymbol{r}_{\mathrm{e},\text{1}}^{T},\dots,\boldsymbol{r}_{\mathrm{e},J}^{T},\varrho_{\mathrm{e},\text{1}}^{\mathrm{R}},\varrho_{\mathrm{e},\text{1}}^{\mathrm{I}},\dots,\varrho_{\mathrm{e},J}^{\mathrm{R}},\varrho_{\mathrm{e},J}^{\mathrm{I}}\right]^{T}$, where $\varrho_{\mathrm{e},j}^{\mathrm{R}}=\mathrm{Re}\left\{ \varrho_{\mathrm{e},j}\right\} $ and $\varrho_{\mathrm{e},j}^{\mathrm{I}}=\mathrm{Im}\left\{ \varrho_{\mathrm{e},j}\right\} $. At snapshot $\ell$, the transmitted waveform $\mathbf{x}\left[\ell\right]=\mathbf{W}_{\mathrm{s}}\boldsymbol{s}_{\mathrm{s}}\left[\ell\right]+\mathbf{W}_{\mathrm{c}}\boldsymbol{s}_{\mathrm{c}}\left[\ell\right]$ for $\ell\in\left\{\text{1},\ldots,L\right\}$ is known at the BS. The corresponding echo vector is written as $\boldsymbol{y}_{\mathrm{s}}\left[\ell\right]=\boldsymbol{\mu}\left(\boldsymbol{\zeta}\mid\mathbf{x}\left[\ell\right]\right)+\boldsymbol{\omega}\left[\ell\right]$, where the conditional mean is
\begin{equation}
\boldsymbol{\mu}\left(\boldsymbol{\zeta}\mid\mathbf{x}\left[\ell\right]\right)=\sum_{j=\text{1}}^{J}\varrho_{\mathrm{e},j}\bar{\boldsymbol{G}}\boldsymbol{\Theta}_{\mathrm{s}}\boldsymbol{h}_{\mathrm{e},j}\boldsymbol{h}_{\mathrm{e},j}^{T}\boldsymbol{\Theta}_{\mathrm{c}}\boldsymbol{G}\mathbf{x}\left[\ell\right],
\end{equation}
and the effective disturbance is
\begin{align}
\boldsymbol{\omega}\left[\ell\right]= & \sum_{j=\text{1}}^{J}\varrho_{\mathrm{e},j}\bar{\boldsymbol{G}}\boldsymbol{\Theta}_{\mathrm{s}}\boldsymbol{h}_{\mathrm{e},j}\boldsymbol{h}_{\mathrm{e},j}^{T}\boldsymbol{\Phi}_{\mathrm{a}}^{\mathrm{c}}\boldsymbol{n}_{\mathrm{a}}^{\mathrm{c}}\left[\ell\right]+\nonumber \\
 & \bar{\boldsymbol{G}}\boldsymbol{\Phi}_{\mathrm{a}}^{\mathrm{s}}\boldsymbol{n}_{\mathrm{a}}^{\mathrm{s}}\left[\ell\right]+\boldsymbol{n}_{\mathrm{BR}}\left[\ell\right].
\end{align}
The first term in \(\boldsymbol{\omega}[\ell]\) represents the DL amplifier noise reflected by the targets, which is assumed negligible due to the additional round-trip attenuation and omitted from the CRB evaluation~\cite{r26}. The UL amplifier noise $\boldsymbol{n}_{\mathrm{a}}^{\mathrm{s}}\left[\ell\right]$ and BS disturbance $\boldsymbol{n}_{\mathrm{BR}}\left[\ell\right]$ are assumed mutually independent and independent of the transmitted waveform. In addition, $\mathbf{R}_{\mathrm{BR}}$ is treated as fixed and locally independent of the target parameters within each CPI. Under this noise approximation, the covariance matrix of $\boldsymbol{\omega}\left[\ell\right]$ is given by
\begin{equation}
\mathbf{R}_{\omega}\triangleq\mathbb{E}\left[\boldsymbol{\omega}\left[\ell\right]\boldsymbol{\omega}^{H}\left[\ell\right]\right]=\delta_{\mathrm{a}}^{\text{2}}\bar{\boldsymbol{G}}\boldsymbol{\Phi}_{\mathrm{a}}^{\mathrm{s}}\left(\boldsymbol{\Phi}_{\mathrm{a}}^{\mathrm{s}}\right)^{H}\bar{\boldsymbol{G}}^{H}+\mathbf{R}_{\mathrm{BR}}.
\end{equation}

Assuming independent disturbances across snapshots, the stacked observation $\boldsymbol{y}\triangleq\left[\boldsymbol{y}_{\mathrm{s}}^{T}\left[\text{1}\right],\cdots,\boldsymbol{y}_{\mathrm{s}}^{T}\left[L\right]\right]^{T}\in\mathbb{C}^{LN\times\text{1}}$ follows $\boldsymbol{y}\sim\mathcal{CN}\left(\boldsymbol{\mu}\left(\boldsymbol{\zeta}\right),\mathbf{I}_{L}\otimes\mathbf{R}_{\omega}\right)$. Here, the Gaussian distribution $\boldsymbol{y}$ is conditioned on the known transmitted waveforms, and the mean vector $\boldsymbol{\mu}\left(\boldsymbol{\zeta}\right)$ stacks the conditional means across snapshots. Since the noise covariance $\mathbf{R}_{\omega}$ is locally independent of the target parameters, the $\left(\zeta_{a,i},\zeta_{b,j}\right)$-th entry of the Fisher information matrix (FIM) $\mathbf{F}\in\mathbb{R}^{\text{5}J\times\text{5}J}$ is evaluated via the Slepian-Bangs formula using the conditional mean derivatives at each snapshot as
\begin{equation}
\left[\mathbf{F}\right]_{\zeta_{a,i},\zeta_{b,j}}=\text{2}\sum_{\ell=\text{1}}^{L}\text{Re}\left\{ \left(\frac{\partial\boldsymbol{\mu}}{\partial\zeta_{a,i}}\right)^{H}\mathbf{R}_{\omega}^{-\text{1}}\left(\frac{\partial\boldsymbol{\mu}}{\partial\zeta_{b,j}}\right)\right\},
\label{eq:FIM}
\end{equation}
where $a,b\in\left\{ x_{\mathrm{e}},y_{\mathrm{e}},z_{\mathrm{e}},\varrho_{\mathrm{e}}^{\mathrm{R}},\varrho_{\mathrm{e}}^{\mathrm{I}}\right\}$ and $i,j\in\mathcal{J}$. We partition the FIM into blocks associated with the target positions and reflection coefficients as
\begin{equation}
\mathbf{F}=\begin{bmatrix}\mathbf{F}_{\boldsymbol{r}\boldsymbol{r}} & \mathbf{F}_{\boldsymbol{r}\boldsymbol{\varrho}}\\
\mathbf{F}_{\boldsymbol{\varrho}\boldsymbol{r}} & \mathbf{F}_{\boldsymbol{\varrho}\boldsymbol{\varrho}}
\end{bmatrix},
\end{equation}
where $\mathbf{F}_{\boldsymbol{r}\boldsymbol{r}}\in\mathbb{R}^{\text{3}J\times\text{3}J}$ corresponds to the target coordinates, and $\mathbf{F}_{\boldsymbol{\varrho}\boldsymbol{\varrho}}\in\mathbb{R}^{\text{2}J\times\text{2}J}$ corresponds to the real and imaginary components of the reflection coefficients. The off-diagonal blocks capture the coupling between these parameter groups.

In the near field, target positions affect path attenuation and spherical-wave phase responses. We account for both effects when evaluating the spatial derivatives in \eqref{eq:FIM}. The normalized NLoS realization, distance-independent gains, Rician factors, and molecular absorption coefficient are treated as known and locally fixed during differentiation. For the first spatial coordinate, let $\zeta_{x_\mathrm{e},j}=x_{\mathrm{e},j}$ and define $\mathbf{H}_{\mathrm{e},j}=\varrho_{\mathrm{e},j}\bar{\boldsymbol{G}}\boldsymbol{\Theta}_{\mathrm{s}}\boldsymbol{h}_{\mathrm{e},j}\boldsymbol{h}_{\mathrm{e},j}^{T}\boldsymbol{\Theta}_{\mathrm{c}}\boldsymbol{G}$. The distance derivative is $ \frac{\partial d_{\mathrm{e},m,j}}{\partial \zeta_{x_\mathrm{e},j}}=\frac{\left(\zeta_{x_\mathrm{e},j}-x_{\mathrm{ris},m}\right)}{d_{\mathrm{e},m,j}}$, and the derivative of the conditional mean is
\begin{equation}
\frac{\partial\boldsymbol{\mu}\left(\boldsymbol{\zeta}\mid\mathbf{x}\left[\ell\right]\right)}{\partial\zeta_{x_\mathrm{e},j}}=\frac{\partial\mathbf{H}_{\mathrm{e},j}}{\partial\zeta_{x_\mathrm{e},j}}\mathbf{x}\left[\ell\right]\triangleq\mathbf{H}_{\zeta_{x_\mathrm{e},j}}^{\prime}\mathbf{x}\left[\ell\right],
\end{equation}
where $\mathbf{H}_{\zeta_{x_\mathrm{e},j}}^{\prime}\!\!=\!\!\varrho_{\mathrm{e},j}\bar{\boldsymbol{G}}\boldsymbol{\Theta}_{\mathrm{s}}\left(\frac{\partial\boldsymbol{h}_{\mathrm{e},j}}{\partial\zeta_{x_\mathrm{e},j}}\boldsymbol{h}_{\mathrm{e},j}^{T}\!+\!\boldsymbol{h}_{\mathrm{e},j}\!\left(\frac{\partial\boldsymbol{h}_{\mathrm{e},j}}{\partial\zeta_{x_\mathrm{e},j}}\right)^{T}\right)\boldsymbol{\Theta}_{\mathrm{c}}\boldsymbol{G}$. The corresponding channel derivative is
\begin{equation}
\frac{\partial\boldsymbol{h}_{\mathrm{e},j}}{\partial\zeta_{x_\mathrm{e},j}}\!\!=\!\! \sqrt{\beta_{\mathrm{e},j}}\!\left(\!\frac{\partial\boldsymbol{\Upsilon}_{\mathrm{e},j}}{\partial\zeta_{x_\mathrm{e},j}}\!\odot\!\boldsymbol{h}_{\mathrm{e},j}^{\mathrm{mix}}\!+\!\sqrt{\frac{\kappa_{\mathrm{e},j}}{\text{1}\!+\!\kappa_{\mathrm{e},j}}}\boldsymbol{\Upsilon}_{\mathrm{e},j}\!\odot\!\frac{\partial\boldsymbol{h}_{\mathrm{e},j}^{\mathrm{LoS}}}{\partial\zeta_{x_\mathrm{e},j}}\right),
\end{equation}
with $\boldsymbol{h}_{\mathrm{e},j}^{\mathrm{mix}}\!=\!\left(\!\sqrt{\frac{\kappa_{\mathrm{e},j}}{\text{1}\!+\!\kappa_{\mathrm{e},j}}}\boldsymbol{h}_{\mathrm{e},j}^{\mathrm{LoS}}\!+\!\sqrt{\frac{\text{1}}{\text{1}\!+\!\kappa_{\mathrm{e},j}}}\boldsymbol{h}_{\mathrm{e},j}^{\mathrm{NLoS}}\right)$. The spatial derivatives of the attenuation and LoS response entries are given by
\begin{equation}
\frac{\partial\left[\boldsymbol{\Upsilon}_{\mathrm{e},j}\right]_{m}}{\partial\zeta_{x_\mathrm{e},j}}=-\lambda_{c}\frac{\partial d_{\mathrm{e},m,j}}{\partial\zeta_{x_\mathrm{e},j}}\frac{\left(\mu_{\mathrm{abs}}d_{\mathrm{e},m,j}+\text{2}\right)}{\text{8}\pi d_{\mathrm{e},m,j}^{\text{2}}}e^{-\frac{\mu_{\mathrm{abs}}d_{\mathrm{e},m,j}}{\text{2}}},
\end{equation}
and
\begin{equation}
\frac{\partial\left[\boldsymbol{h}_{\mathrm{e},j}^{\mathrm{LoS}}\right]_{m}}{\partial\zeta_{x_\mathrm{e},j}}=-\mathrm{j}\frac{\text{2}\pi}{\lambda_{c}}\frac{\partial d_{\mathrm{e},m,j}}{\partial\zeta_{x_\mathrm{e},j}}e^{-\mathrm{j}\frac{\text{2}\pi}{\lambda_{c}}d_{\mathrm{e},m,j}}.
\end{equation}
The derivatives with respect to $y_{\mathrm{e},j}$ and $z_{\mathrm{e},j}$ are obtained similarly. Since the echo response is linear in each target reflection coefficient, the derivatives with respect to its real and imaginary components are
\begin{equation}
\mathbf{H}_{\zeta_{\varrho_{\mathrm{e}}^{\mathrm{R}},j}}^{\prime}=\frac{\partial\mathbf{H}_{\mathrm{e},j}}{\partial\zeta_{\varrho_{\mathrm{e}}^{\mathrm{R}},j}}=\bar{\boldsymbol{G}}\boldsymbol{\Theta}_{\mathrm{s}}\boldsymbol{h}_{\mathrm{e},j}\boldsymbol{h}_{\mathrm{e},j}^{T}\boldsymbol{\Theta}_{\mathrm{c}}\boldsymbol{G},
\end{equation}
\begin{equation}
\mathbf{H}_{\zeta_{\varrho_{\mathrm{e}}^{\mathrm{I}},j}}^{\prime}=\frac{\partial\mathbf{H}_{\mathrm{e},j}}{\partial\zeta_{\varrho_{\mathrm{e}}^{\mathrm{I}},j}}=\mathrm{j}\bar{\boldsymbol{G}}\boldsymbol{\Theta}_{\mathrm{s}}\boldsymbol{h}_{\mathrm{e},j}\boldsymbol{h}_{\mathrm{e},j}^{T}\boldsymbol{\Theta}_{\mathrm{c}}\boldsymbol{G}.
\end{equation}
For a unified representation, define
\begin{equation}
\mathbf{A}_{\zeta_{a,i}}=\begin{cases}
\varrho_{\mathrm{e},i}\!\left(\!\frac{\partial\boldsymbol{h}_{\mathrm{e},i}}{\partial\zeta_{a,i}}\boldsymbol{h}_{\mathrm{e},i}^{T}\!+\!\boldsymbol{h}_{\mathrm{e},i}\left(\frac{\partial\boldsymbol{h}_{\mathrm{e},i}}{\partial\zeta_{a,i}}\right)^{T}\right),\!\!\! & a\in\left\{ x_{\mathrm{e}},y_{\mathrm{e}},z_{\mathrm{e}}\right\} ,\\
\boldsymbol{h}_{\mathrm{e},i}\boldsymbol{h}_{\mathrm{e},i}^{T},\! & a=\varrho_{\mathrm{e}}^{\mathrm{R}},\\
\mathrm{j}\boldsymbol{h}_{\mathrm{e},i}\boldsymbol{h}_{\mathrm{e},i}^{T},\! & a=\varrho_{\mathrm{e}}^{\mathrm{I}},
\label{eq:CRB_unified representation}
\end{cases}
\end{equation}
which gives $\mathbf{H}_{\zeta_{a,i}}^{\prime}=\bar{\boldsymbol{G}}\boldsymbol{\Theta}_{\mathrm{s}}\mathbf{A}_{\zeta_{a,i}}\boldsymbol{\Theta}_{\mathrm{c}}\boldsymbol{G}$.

For an ergodic transmitted waveform and sufficiently large $L$, the sample covariance is approximated by its statistical covariance, i.e.,
$\frac{\text{1}}{L}\sum_{\ell=\text{1}}^{L}\mathbf{x}\left[\ell\right]\mathbf{x}^{H}\left[\ell\right]\approx\mathbf{R}_{\mathrm{x}}$, where $\mathbf{R}_{\mathrm{x}}\triangleq
\mathbb{E}\left[\mathbf{x}\left[\ell\right]\mathbf{x}^{H}\left[\ell\right]\right]
=
\mathbf{W}_{\mathrm{c}}\mathbf{W}_{\mathrm{c}}^{H}
+
\mathbf{W}_{\mathrm{s}}\mathbf{W}_{\mathrm{s}}^{H}
$. Substituting this covariance approximation and the derived response derivatives into \eqref{eq:FIM}, the $\left(\zeta_{a,i},\zeta_{b,j}\right)$-th FIM entry is approximated as
\begin{equation}
\left[\mathbf{F}\right]_{\zeta_{a,i},\zeta_{b,j}}=\text{2}L\mathrm{Re}\left\{ \text{Tr}\left(\left(\mathbf{H}_{\zeta_{a,i}}^{\prime}\right)^{H}\mathbf{R}_{\omega}^{-\text{1}}\mathbf{H}_{\zeta_{b,j}}^{\prime}\mathbf{R}_{\mathrm{x}}\right)\right\}.
\label{eq:FIM_2}
\end{equation}
For a locally identifiable state with a positive-definite FIM, we treat the complex reflection coefficients as nuisance parameters in position estimation. By eliminating these nuisance parameters through the Schur complement, the equivalent 3D position CRB matrix 
$\mathrm{CRB}\left(\boldsymbol{r}\right)=\left(\mathbf{F}_{\boldsymbol{r}\boldsymbol{r}}-\mathbf{F}_{\boldsymbol{r}\boldsymbol{\varrho}}\mathbf{F}_{\boldsymbol{\varrho}\boldsymbol{\varrho}}^{-\text{1}}\mathbf{F}_{\boldsymbol{\varrho}\boldsymbol{r}}\right)^{-\text{1}}$.
For the $j$-th target, $\mathrm{CRB}(\boldsymbol{r}_{\mathrm{e},j})\in\mathbb{R}^{\text{3}\times\text{3}}$ is the $j$-th $\text{3}\times\text{3}$ diagonal block of $\mathrm{CRB}(\boldsymbol{r})$. Under the adopted observation and covariance approximations, $\mathrm{Tr}\left(\mathrm{CRB}(\boldsymbol{r}_{\mathrm{e},j})\right)$ provides a lower bound on the mean squared error achieved by any unbiased estimator of the target's 3D position.

%%%%%%%%%%%%%%%%%%%%%%%%%%%%%%%%%%%%%%%%%%%%
\vspace{-4mm}
\subsection{Problem Formulation}

In this paper, we maximize the system SEE to balance secrecy performance and total power consumption subject to user rate, CRB-based localization accuracy, and power consumption. The SEE is defined as the ratio of the achievable sum secrecy rate to the total system power consumption. The achievable sum secrecy rate is given by $R_{\mathrm{sec}}=\sum_{k=\text{1}}^{K}\left[\log_{\text{2}}\left(\text{1}+\gamma_{\mathrm{u},k}\right)-\underset{j\in\mathcal{J}}{\max}\log_{\text{2}}\left(\text{1}+\gamma_{\mathrm{e},j,k}\right)\right]^{+}$.
Assuming unit power-amplifier efficiency at the BS and HRIS \cite{r24}, the total system power consumption is given by
\begin{equation}
P_{\mathrm{tot}}=\mathrm{Tr}\left(\mathbf{R}_{\mathrm{x}}\right)+P_{\mathrm{f}}+P_{\mathrm{ris}},
\label{eq:total_power}
\end{equation}
where $\mathrm{Tr}\left(\mathbf{R}_{\mathrm{x}}\right)$ is the BS transmit power, $P_{\mathrm{f}}$ denotes the configuration-independent static hardware power consumption, and $P_{\mathrm{ris}}$ is the power consumed by the HRIS. The HRIS power consumption comprises mode-dependent circuit power and amplified output power from both branches, i.e.,
\begin{equation}
P_{\mathrm{ris}}=\left\Vert \boldsymbol{e}\right\Vert _{\text{1}}P_{\mathrm{a}}+\left(M-\left\Vert \boldsymbol{e}\right\Vert _{\text{1}}\right)P_{\mathrm{p}}+P_{\mathrm{c}}+P_{\mathrm{s}},
\label{eq:hris_power}
\end{equation}
where $P_{\mathrm{a}}$ and $P_{\mathrm{p}}$ denote the static circuit powers per element in active and passive modes, respectively. The amplified output powers of the DL and UL branches are $P_{\mathrm{c}}= \left\Vert \boldsymbol{\Phi}_{\mathrm{a}}^{\mathrm{c}}\boldsymbol{G}\mathbf{W}_{\mathrm{s}}\right\Vert _{F}^{\text{2}}
+\left\Vert \boldsymbol{\Phi}_{\mathrm{a}}^{\mathrm{c}}\boldsymbol{G}\mathbf{W}_{\mathrm{c}}\right\Vert _{F}^{\text{2}}+\delta_{\mathrm{a}}^{\text{2}}\left\Vert \boldsymbol{\Phi}_{\mathrm{a}}^{\mathrm{c}}\right\Vert _{F}^{\text{2}}$ and $P_{\mathrm{s}}=P_{\mathrm{echo}}+\delta_{\mathrm{a}}^{\text{2}}\left\Vert \boldsymbol{\Phi}_{\mathrm{a}}^{\mathrm{s}}\right\Vert _{F}^{\text{2}}$, respectively.
For the long-term power model, we assume that the cross-target echo correlation terms average to zero over time. Under this assumption, the echo power amplified by the UL active branch is expressed as 
\begin{equation}
P_{\mathrm{echo}}\!=\!\sum_{j=\text{1}}^{J}\left|\varrho_{\mathrm{e},j}\right|^{\text{2}}\left(\left\Vert \boldsymbol{\Lambda}_{\mathrm{e},j}\boldsymbol{G}\mathbf{W}_{\mathrm{s}}\right\Vert _{F}^{\text{2}}\!+\!\left\Vert \boldsymbol{\Lambda}_{\mathrm{e},j}\boldsymbol{G}\mathbf{W}_{\mathrm{c}}\right\Vert _{F}^{\text{2}}\!+\!\delta_{\mathrm{a}}^{\text{2}}\left\Vert \bar{\boldsymbol{\Lambda}}_{\mathrm{e},j}\right\Vert _{F}^{\text{2}}\right), 
\end{equation}
where $\boldsymbol{\Lambda}_{\mathrm{e},j}=\boldsymbol{\Phi}_{\mathrm{a}}^{\mathrm{s}}\boldsymbol{h}_{\mathrm{e},j}\boldsymbol{h}_{\mathrm{e},j}^{T}\boldsymbol{\Theta}_{\mathrm{c}}$ and $\bar{\boldsymbol{\Lambda}}_{\mathrm{e},j}=\boldsymbol{\Phi}_{\mathrm{a}}^{\mathrm{s}}\boldsymbol{h}_{\mathrm{e},j}\boldsymbol{h}_{\mathrm{e},j}^{T}\boldsymbol{\Phi}_{\mathrm{a}}^{\mathrm{c}}$.

We collect the continuous design variables in the set $\mathcal{X}=\left\{\mathbf{W}_{\mathrm{c}},\mathbf{W}_{\mathrm{s}},\boldsymbol{\alpha},\boldsymbol{\rho},\boldsymbol{\phi}_{\mathrm{a}}^{\mathrm{c}},\boldsymbol{\phi}_{\mathrm{a}}^{\mathrm{s}},\boldsymbol{\phi}_{\mathrm{p}}\right\}$. Then, the joint active-element configuration and transmission design problem for SEE maximization is formulated as
\begin{subequations}
\label{eq:P0}
\begin{eqnarray}
\mathrm{P\text{0}}: & \underset{\mathcal{X},\boldsymbol{e}}{\max} & \frac{R_{\mathrm{sec}}}{P_{\mathrm{tot}}}
\label{eq:P0_obj}\\
& \mathrm{s.t.} & \mathrm{Tr}\left(\mathbf{R}_{\mathrm{x}}\right)\leq P_{\max},
\label{eq:P0_bs_power}\\
&& P_{\mathrm{ris}}\leq P_{\mathrm{ris},\max},
\label{eq:P0_hris_power}\\
&& R_{\mathrm{u},k}\geq R_{\mathrm{u},\min},\forall k\in\mathcal{K},
\label{eq:P0_user_rate}\\
&& \mathrm{Tr}\left(\mathrm{CRB}\left(\boldsymbol{r}_{\mathrm{e},j}\right)\right)\leq\epsilon_{\max}^{\text{2}},
\forall j\in\mathcal{J},
\label{eq:P0_crb}\\
&& \text{1}\leq\alpha_{m}\leq\alpha_{\max},\forall m\in\mathcal{M},
\label{eq:P0_gain}\\
&& \text{0}\leq\rho_{m}\leq\text{1},\forall m\in\mathcal{M}, 
\label{eq:P0_split}\\
&& \left|\left[\boldsymbol{\phi}_{\mathrm{a}}^{\mathrm{c}}\right]_{m}\right|=\text{1},
\left|\left[\boldsymbol{\phi}_{\mathrm{a}}^{\mathrm{s}}\right]_{m}\right|=\text{1},
\forall m\in\mathcal{M},
\label{eq:P0_active_phase}\\
&& \left|\left[\boldsymbol{\phi}_{\mathrm{p}}\right]_{m}\right|=\text{1},
\forall m\in\mathcal{M},
\label{eq:P0_passive_phase}\\
&& \boldsymbol{e}\in\left\{\text{0},\text{1}\right\}^{M}.
\label{eq:P0_selection}
\end{eqnarray}
\end{subequations}
In problem $\mathrm{P\text{0}}$, constraints \eqref{eq:P0_bs_power} and \eqref{eq:P0_hris_power} enforce the BS transmit power budget and the HRIS total power budget, respectively. Constraint \eqref{eq:P0_user_rate} guarantees the minimum rate requirement for each legitimate user, while \eqref{eq:P0_crb} imposes a CRB-based localization accuracy requirement, with $\epsilon_{\max}$ denoting the prescribed CRB-based localization root-mean-square error (RMSE) threshold. Constraints \eqref{eq:P0_gain} and \eqref{eq:P0_split} restrict the amplification gains and power-splitting ratios, respectively. Constraints \eqref{eq:P0_active_phase} and \eqref{eq:P0_passive_phase} impose unit-modulus restrictions on the phase shifts, and \eqref{eq:P0_selection} defines the active-element selection indicators.
Problem $\mathrm{P\text{0}}$ is a mixed-integer non-convex fractional program involving a fractional objective, the binary selection vector $\boldsymbol{e}$, and continuous transmission variables highly coupled through the secrecy-rate, localization, and power expressions. To tackle this challenging problem, an AO framework is developed in the next section.

%%%%%%%%%%%%%%%%%%%%%%%%%%%%%%%%%%%%%%%%%%%%
\vspace{-2mm}
\section{Proposed Algorithm}

In this section, we present an AO framework to solve problem $\mathrm{P\text{0}}$. We first reformulate the SEE objective and apply SCA to the coupled non-convex constraints. Then, the resulting subproblems are solved alternately to update the continuous transmission variables in $\mathcal{X}$ and the active-element configuration. Specifically, the active and passive HRIS phase shifts are optimized directly in vector form through PCCP. The active-element configuration is updated using a full-aperture reweighted group-sparse selection method.

\vspace{-3mm}

\subsection{Problem Transformation}

In this subsection, we first reformulate the SEE objective in \eqref{eq:P0_obj} with auxiliary variables $\psi\geq\text{0}$ and $\eta>\text{0}$. Problem $\mathrm{P\text{0}}$ is equivalently written as
\begin{subequations}
\label{eq:P1}
\begin{eqnarray}
\mathrm{P\text{1}}: & \underset{\mathcal{X},\boldsymbol{e},\psi,\eta}{\max} & \psi
\label{eq:P1_obj}\\
& \mathrm{s.t.} & R_{\mathrm{sec}}\geq\psi\eta,
\label{eq:P1_rate}\\
&& P_{\mathrm{tot}}\leq\eta,
\label{eq:P1_power}\\
&& \eqref{eq:P0_bs_power}-\eqref{eq:P0_selection},
\label{eq:P1_remaining}
\end{eqnarray}
\end{subequations}
where $\psi=R_{\mathrm{sec}}/P_{\mathrm{tot}}$ at a positive-SEE optimum. From a positive-SEE feasible point, SCA upper bounds the product at iteration $t$ by
$\psi\eta\leq\mathcal{G}^{(t)}(\psi,\eta)
\triangleq
\frac{\xi^{(t)}}{\text{2}}\psi^{\text{2}}
+\frac{\eta^{\text{2}}}{\text{2}\xi^{(t)}}$,
where $\xi^{(t)}=\eta^{(t)}/\psi^{(t)}>\text{0}$. This convex bound is tight at the current iterate. With auxiliary rate bounds
$\boldsymbol{v}\triangleq
\left\{
v_{\mathrm{u},k},v_{\mathrm{e},k}\geq\text{0}
\mid k\in\mathcal{K}
\right\}$, using this bound and requiring nonnegative secrecy rate at each user gives a conservative restriction of $\mathrm{P\text{1}}$ as follows 
\begin{subequations}
\label{eq:P2}
\begin{eqnarray}
\mathrm{P\text{2}}: & \underset{\mathcal{X},\boldsymbol{e},\psi,\eta,\boldsymbol{v}}{\max} & \psi
\label{eq:P2_obj}\\
& \mathrm{s.t.} &
R_{\mathrm{u},k}\geq v_{\mathrm{u},k},
\forall k\in\mathcal{K},
\label{eq:P2_user_rate}\\
&&
R_{\mathrm{e},j,k}\leq v_{\mathrm{e},k},
\forall j\in\mathcal{J},\forall k\in\mathcal{K},
\label{eq:P2_eve_rate}\\
&&
v_{\mathrm{u},k}\geq R_{\mathrm{u},\min},
\forall k\in\mathcal{K},
\label{eq:P2_qos}\\
&&
\sum_{k=\text{1}}^{K}
\left(v_{\mathrm{u},k}-v_{\mathrm{e},k}\right)
\geq
\mathcal{G}^{(t)}(\psi,\eta),
\label{eq:P2_see}\\
&&
v_{\mathrm{u},k}\geq v_{\mathrm{e},k},
\forall k\in\mathcal{K},
\label{eq:P2_positive}\\
&&
\eqref{eq:P0_bs_power},
\eqref{eq:P0_hris_power},
\eqref{eq:P0_crb}-\eqref{eq:P0_selection},
\eqref{eq:P1_power}.
\label{eq:P2_remaining}
\end{eqnarray}
\end{subequations}

To facilitate the convex approximation of
\eqref{eq:P2_user_rate} and \eqref{eq:P2_eve_rate},
we define the signal and interference-plus-noise powers
for receiver $(p,q)$ and stream $k$ as
$D_{p,q,k}
=|\boldsymbol{h}_{p,q}^{T}\boldsymbol{\Theta}_{\mathrm{c}}
\boldsymbol{G}\boldsymbol{w}_{\mathrm{c},k}|^{\text{2}}$
and
$I_{p,q,k}
=c_{p,q,k}
+\delta_{\mathrm{a}}^{\text{2}}
\|\boldsymbol{h}_{p,q}^{T}
\boldsymbol{\Phi}_{\mathrm{a}}^{\mathrm{c}}\|^{\text{2}}
+\delta_{p,q}^{\text{2}}$,
respectively. By introducing $\boldsymbol{\varpi}\triangleq\left\{\varpi_{\mathrm{u},k},\varpi_{\mathrm{e},j,k}\geq\text{0}\mid k\in\mathcal{K},j\in\mathcal{J}\right\}$,  \eqref{eq:P2_user_rate} and \eqref{eq:P2_eve_rate} are equivalently split into
\begin{subequations}
\label{eq:sinr_split}
\begin{alignat}{2}
& D_{\mathrm{u},k}\geq
\varpi_{\mathrm{u},k}
\left(\text{2}^{v_{\mathrm{u},k}}-\text{1}\right),
&& D_{\mathrm{e},j,k}\leq
\varpi_{\mathrm{e},j,k}
\left(\text{2}^{v_{\mathrm{e},k}}-\text{1}\right),
\label{eq:signal_split}\\
& I_{\mathrm{u},k}\leq
\varpi_{\mathrm{u},k},
&&I_{\mathrm{e},j,k}\geq
\varpi_{\mathrm{e},j,k}.
\label{eq:interference_split}
\end{alignat}
\end{subequations}
Then, we use SCA to upper-bound the user-side product and lower-bound the Eve-side product. For the user-side product,  define
$g_{\mathrm{u},k}
=\varpi_{\mathrm{u},k}\left(\text{2}^{v_{\mathrm{u},k}}-\text{1}\right)$.
Given a positive coefficient $\xi_{\mathrm{u},k}^{(t)}$, the convex upper bound is
$g_{\mathrm{u},k}
\leq
\frac{\xi_{\mathrm{u},k}^{(t)}}{\text{2}}\varpi_{\mathrm{u},k}^{\text{2}}
+
\frac{\text{1}}{\text{2}\xi_{\mathrm{u},k}^{(t)}}
\left(\text{2}^{v_{\mathrm{u},k}}-\text{1}\right)^{\text{2}}
\triangleq
g_{\mathrm{u},k}^{\mathrm{ub},(t)}$,
where
$\xi_{\mathrm{u},k}^{(t)}
=
\left(\text{2}^{v_{\mathrm{u},k}^{(t)}}-\text{1}\right)/
\varpi_{\mathrm{u},k}^{(t)}$.
For the Eve-side product, define $l_{\mathrm{e},k}=\text{2}^{v_{\mathrm{e},k}}-\text{1}$.
Its affine Taylor lower bound is
$\hat{l}_{\mathrm{e},k}^{(t)}
=
\text{2}^{v_{\mathrm{e},k}^{(t)}}-\text{1}
+
\text{2}^{v_{\mathrm{e},k}^{(t)}}\ln \text{2}
\left(v_{\mathrm{e},k}-v_{\mathrm{e},k}^{(t)}\right)$.
Since $\hat{l}_{\mathrm{e},k}^{(t)}\leq l_{\mathrm{e},k}$ and $\varpi_{\mathrm{e},j,k}\geq\text{0}$, a concave lower bound on
$\varpi_{\mathrm{e},j,k}l_{\mathrm{e},k}$ is
\begin{align}
\varpi_{\mathrm{e},j,k}l_{\mathrm{e},k}
\geq&
\frac{\text{1}}{\text{4}}
\left[
\text{2}\xi_{\mathrm{e},j,k}^{(t)}
\left(
\varpi_{\mathrm{e},j,k}
+
\hat{l}_{\mathrm{e},k}^{(t)}
\left(v_{\mathrm{e},k}\right)
\right)
-
\left(\xi_{\mathrm{e},j,k}^{(t)}\right)^{\text{2}}
\right.
\nonumber\\
&\left.
-
\left(
\varpi_{\mathrm{e},j,k}
-
\hat{l}_{\mathrm{e},k}^{(t)}
\left(v_{\mathrm{e},k}\right)
\right)^{\text{2}}
\right]
\triangleq
g_{\mathrm{e},j,k}^{\mathrm{lb},(t)},
\label{eq:eve_product_lb}
\end{align}
where
$\xi_{\mathrm{e},j,k}^{(t)}
=
\varpi_{\mathrm{e},j,k}^{(t)}
+
\left(\text{2}^{v_{\mathrm{e},k}^{(t)}}-\text{1}\right)$. Accordingly, at iteration $t$, \eqref{eq:signal_split} is conservatively approximated by
\begin{equation}
D_{\mathrm{u},k}
\geq
g_{\mathrm{u},k}^{\mathrm{ub},(t)},\quad
D_{\mathrm{e},j,k}
\leq
g_{\mathrm{e},j,k}^{\mathrm{lb},(t)},
\quad
\forall j\in\mathcal{J},\forall k\in\mathcal{K}.
\label{eq:signal_split1}
\end{equation}

To reformulate the non-convex localization constraint \eqref{eq:P0_crb}, we define
$\mathbf{J}_{j}
=
\left[
\mathbf{\text{0}}_{\text{3}\times\text{3}(j-\text{1})},
\mathbf{I}_{\text{3}},
\mathbf{\text{0}}_{\text{3}\times\text{3}(J-j)},
\mathbf{\text{0}}_{\text{3}\times\text{2}J}
\right]
\in\mathbb{R}^{\text{3}\times\text{5}J}$
to select the position coordinates of target $j$. For $\mathbf{F}\succ\boldsymbol{\text{0}}$, the CRB is
$\mathrm{CRB}\left(\boldsymbol{r}_{\mathrm{e},j}\right)
=
\mathbf{J}_{j}\mathbf{F}^{-\text{1}}\mathbf{J}_{j}^{T}$.
By introducing $\mathbf{C}_{j}\in\mathbb{S}_{+}^{\text{3}}$, the equivalent Schur complement form of \eqref{eq:P0_crb} is given by
\begin{subequations}
\label{eq:crb_schur}
\begin{align}
&\begin{bmatrix}
\mathbf{F} & \mathbf{J}_{j}^{T}\\
\mathbf{J}_{j} & \mathbf{C}_{j}
\end{bmatrix}
\succeq\boldsymbol{\text{0}},
\forall j\in\mathcal{J},
\label{eq:crb_schur_lmi}\\
&\mathrm{Tr}\left(\mathbf{C}_{j}\right)
\leq\epsilon_{\max}^{\text{2}},
\forall j\in\mathcal{J}.
\label{eq:crb_trace_aux}
\end{align}
\end{subequations}

With auxiliary variables
$\boldsymbol{\Delta}\triangleq
\left\{
\psi,\eta,\boldsymbol{v},\boldsymbol{\varpi},
\left\{\mathbf{C}_{j}\right\}
\right\}$, the resulting surrogate problem is given by
\begin{subequations}
\label{eq:P3}
\begin{eqnarray}
\mathrm{P\text{3}}: & \underset{\mathcal{X},\boldsymbol{e},\boldsymbol{\Delta}}{\max} & \psi
\label{eq:P3_obj}\\
&\mathrm{s.t.}&
\eqref{eq:P0_bs_power},\eqref{eq:P0_hris_power},
\eqref{eq:P0_gain}-\eqref{eq:P0_selection},
\eqref{eq:P1_power},
\nonumber\\
&&
\eqref{eq:P2_qos}-\eqref{eq:P2_positive},
\eqref{eq:interference_split},
\eqref{eq:signal_split1},\eqref{eq:crb_schur}.
\label{eq:P3_1}
\end{eqnarray}
\end{subequations}
Problem $\mathrm{P\text{3}}$ remains non-convex because of variable coupling and binary active-element selection. The corresponding AO updates are detailed in the following.

%%%%%%%%%%%%%%%%%%%%%%%%%%%%%%%%%%%%%%%%%%%%
\vspace{-2mm}
\subsection{BS Beamforming Optimization}

With all HRIS variables fixed, the BS beamformers  are updated by solving
\begin{subequations}
\label{eq:P4}
\begin{eqnarray}
\mathrm{P\text{4}}: &
\underset{\mathbf{W}_{\mathrm{c}},\mathbf{W}_{\mathrm{s}},\boldsymbol{\Delta}}{\max}
& \psi
\label{eq:P4_obj}\\
& \mathrm{s.t.} &
\eqref{eq:P0_bs_power},\eqref{eq:P0_hris_power},
\eqref{eq:P1_power},\eqref{eq:P2_qos}-\\
&&
\eqref{eq:P2_positive},\eqref{eq:interference_split},
\eqref{eq:signal_split1},\eqref{eq:crb_schur}.
\end{eqnarray}
\end{subequations}

Since the sensing beamformer $\mathbf{W}_{\mathrm{s}}$ enters the rate, power, and FIM expressions only through its covariance, we replace it with the total transmit covariance $\mathbf{R}_{\mathrm{x}}$ while retaining $\mathbf{W}_{\mathrm{c}}$ as an optimization variable. In these variables, the BS power constraint \eqref{eq:P0_bs_power} becomes
$\mathrm{Tr}(\mathbf{R}_{\mathrm{x}})\leq P_{\max}$.
The corresponding sensing covariance is
$\mathbf{S}=\mathbf{R}_{\mathrm{x}}-\mathbf{W}_{\mathrm{c}}\mathbf{W}_{\mathrm{c}}^{H}$, which can be equivalently enforced through the Schur complement as
\begin{equation}
\begin{bmatrix}
\mathbf{R}_{\mathrm{x}} & \mathbf{W}_{\mathrm{c}}\\
\mathbf{W}_{\mathrm{c}}^{H} & \mathbf{I}_{K}
\end{bmatrix}
\succeq\boldsymbol{\text{0}}.
\label{eq:beam_covariance_lmi}
\end{equation}

Then, the HRIS and total power constraints \eqref{eq:P0_hris_power} and \eqref{eq:P1_power} take the affine forms as follows
\begin{subequations}
\label{eq:beam_power}
\begin{align}
&\mathrm{Tr}\left(
\mathbf{\Gamma}_{\mathrm{w}}\mathbf{R}_{\mathrm{x}}
\right)+c_{\mathrm{w}}
\leq P_{\mathrm{ris},\max},
\label{eq:beam_hris_power}
\displaybreak[1]\\
&\mathrm{Tr}\left[
\left(\mathbf{I}_{N}+\mathbf{\Gamma}_{\mathrm{w}}\right)
\mathbf{R}_{\mathrm{x}}
\right]+c_{\mathrm{w}}+P_{\mathrm{f}}
\leq\eta,
\label{eq:beam_total_power}
\end{align}
\end{subequations}
where $\mathbf{\Gamma}_{\mathrm{w}}
=\boldsymbol{G}^{H}\!\left(
(\boldsymbol{\Phi}_{\mathrm{a}}^{\mathrm{c}})^{H}
\boldsymbol{\Phi}_{\mathrm{a}}^{\mathrm{c}}
+\sum_{j=\text{1}}^{J}|\varrho_{\mathrm{e},j}|^{\text{2}}
\boldsymbol{\Lambda}_{\mathrm{e},j}^{H}
\boldsymbol{\Lambda}_{\mathrm{e},j}
\right)\boldsymbol{G}$ and $c_{\mathrm{w}}\!=\!\left\Vert \boldsymbol{e}\right\Vert _{\text{1}}P_{\mathrm{a}}+\left(M-\left\Vert \boldsymbol{e}\right\Vert _{\text{1}}\right)P_{\mathrm{p}}+\delta_{\mathrm{a}}^{\text{2}}\left(\left\Vert \boldsymbol{\Phi}_{\mathrm{a}}^{\mathrm{c}}\right\Vert _{F}^{\text{2}}+\left\Vert \boldsymbol{\Phi}_{\mathrm{a}}^{\mathrm{s}}\right\Vert _{F}^{\text{2}}+\right.$ $\left.\sum_{j=\text{1}}^{J}\left|\varrho_{\mathrm{e},j}\right|^{\text{2}}\left\Vert \bar{\boldsymbol{\Lambda}}_{\mathrm{e},j}\right\Vert _{F}^{\text{2}}\right).$

Next, we address the rate constraints \eqref{eq:interference_split} and \eqref{eq:signal_split1}. For receiver $(p,q)$ and stream $k$, the signal and interference-plus-noise powers are expressed  as
$D_{p,q,k}=\boldsymbol{w}_{\mathrm{c},k}^{H}\mathbf{Y}_{p,q}
\boldsymbol{w}_{\mathrm{c},k}$
and
$I_{p,q,k}=\mathrm{Tr}(\mathbf{Y}_{p,q}\mathbf{R}_{\mathrm{x}})
-D_{p,q,k}+\bar{\delta}_{p,q}$,
respectively, where
$\mathbf{Y}_{p,q}=\boldsymbol{G}^{H}\boldsymbol{\Theta}_{\mathrm{c}}^{H}
\boldsymbol{h}_{p,q}^{\ast}\boldsymbol{h}_{p,q}^{T}
\boldsymbol{\Theta}_{\mathrm{c}}\boldsymbol{G}$
and
$\bar{\delta}_{p,q}=\delta_{\mathrm{a}}^{\text{2}}
\|\boldsymbol{h}_{p,q}^{T}\boldsymbol{\Phi}_{\mathrm{a}}^{\mathrm{c}}\|_{\text{2}}^{\text{2}}
+\delta_{p,q}^{\text{2}}$.
For each user, we lower-bound the convex quadratic term $D_{\mathrm{u},k}$ by its first-order Taylor approximation at $\boldsymbol{w}_{\mathrm{c},k}^{(t)}$, which is given by
\begin{equation}
\underline{D}_{\mathrm{u},k}^{\mathrm{w},(t)}\left(\boldsymbol{w}_{\mathrm{c},k}\right)
\!=\!
\text{2}\mathrm{Re}\!\left\{
\boldsymbol{w}_{\mathrm{c},k}^{(t),H}
\mathbf{Y}_{\mathrm{u},k}\boldsymbol{w}_{\mathrm{c},k}
\right\}
\!-\!
\boldsymbol{w}_{\mathrm{c},k}^{(t),H}
\mathbf{Y}_{\mathrm{u},k}
\boldsymbol{w}_{\mathrm{c},k}^{(t)}.
\label{eq:beam_user_signal_lb}
\end{equation}
Substituting this lower bound into $I_{\mathrm{u},k}$ gives the affine upper bound
$\overline{I}_{\mathrm{u},k}^{\mathrm{w},(t)}
=\mathrm{Tr}(\mathbf{Y}_{\mathrm{u},k}\mathbf{R}_{\mathrm{x}})
-\underline{D}_{\mathrm{u},k}^{\mathrm{w},(t)}
+\bar{\delta}_{\mathrm{u},k}$.
For each Eve, $D_{\mathrm{e},j,k}$ is convex in $\boldsymbol{w}_{\mathrm{c},k}$ and $I_{\mathrm{e},j,k}$ is concave in $(\mathbf{R}_{\mathrm{x}},\boldsymbol{w}_{\mathrm{c},k})$, so both expressions are retained exactly. The rate constraints are conservatively approximated by
\begin{subequations}
\label{eq:sinr_split2}
\begin{alignat}{2}
& \underline{D}_{\mathrm{u},k}^{\mathrm{w},(t)}
\geq g_{\mathrm{u},k}^{\mathrm{ub},(t)},
&&\ D_{\mathrm{e},j,k}
\leq g_{\mathrm{e},j,k}^{\mathrm{lb},(t)},
\forall j\in\mathcal{J},\forall k\in\mathcal{K},
\label{eq:signal_split2}\\
& \overline{I}_{\mathrm{u},k}^{\mathrm{w},(t)}
\leq\varpi_{\mathrm{u},k},
&&\ I_{\mathrm{e},j,k}
\geq\varpi_{\mathrm{e},j,k},
\forall j\in\mathcal{J},\forall k\in\mathcal{K}.
\label{eq:interference_split2}
\end{alignat}
\end{subequations}

Since $\mathbf{R}_{\omega}$ and $\mathbf{H}_{\zeta_{a,i}}^{\prime}$ are fixed in this subproblem, the FIM $\mathbf{F}$ is affine in $\mathbf{R}_{\mathrm{x}}$, making \eqref{eq:crb_schur_lmi}  a linear matrix inequality. Therefore, the convex surrogate of problem $\mathrm{P\text{4}}$ is given by
\begin{subequations}
\label{eq:P5}
\begin{eqnarray}
\mathrm{P\text{5}}:\! &\!
\underset{\mathbf{W}_{\mathrm{c}},\mathbf{R}_{\mathrm{x}},\boldsymbol{\Delta}}{\max}
&
\!\psi
\label{eq:P5_obj}\\
& \!\mathrm{s.t.} &
\!\eqref{eq:P2_qos}\!\!-\!\!\eqref{eq:P2_positive},
\eqref{eq:crb_schur},
\eqref{eq:beam_covariance_lmi},
\eqref{eq:beam_power},
\eqref{eq:sinr_split2},
\label{eq:P5_1}\\
&&
\mathrm{Tr}\left(\mathbf{R}_{\mathrm{x}}\right)
\leq P_{\max}.
\label{eq:P5_bs_power}
\end{eqnarray}
\end{subequations}
After solving $\mathrm{P\text{5}}$ using CVX, we recover $\mathbf{W}_{\mathrm{s}}$ by factoring the optimized sensing covariance $\mathbf{S}$ as $\mathbf{S}=\mathbf{W}_{\mathrm{s}}\mathbf{W}_{\mathrm{s}}^{H}$.

%%%%%%%%%%%%%%%%%%%%%%%%%%%%%%%%%%%%%%%%%%%%
%%%%%%%%%%%%%%%%%%%%%%%%%%%%%%%%%%%%%%%%%%%%
%%%%%%%%%%%%%%%%%%%%%%%%%%%%%%%%%%%%%%%%%%%%
%%%%%%%%%%%%%%%%%%%%%%%%%%%%%%%%%%%%%%%%%%%%
\vspace{-4mm}
\subsection{Amplification and Power-Splitting Optimization}

With the BS beamformers, HRIS phase shifts, and active-element configuration fixed, the amplification gains and power-splitting ratios are updated by solving
\begin{subequations}
\label{eq:P6}
\begin{eqnarray}
\mathrm{P\text{6}}: &
\underset{\boldsymbol{\alpha},\boldsymbol{\rho},\boldsymbol{\Delta}}{\max}
& \psi
\label{eq:P6_obj}\\
&\mathrm{s.t.}&
\eqref{eq:P0_hris_power},\eqref{eq:P0_gain},
\eqref{eq:P0_split},\eqref{eq:P1_power},
\label{eq:P6_1}\\
&&
\eqref{eq:P2_qos}-\eqref{eq:P2_positive},
\eqref{eq:interference_split},
\eqref{eq:signal_split1},\eqref{eq:crb_schur}.
\label{eq:P6_2}
\end{eqnarray}
\end{subequations}
To express the DL and UL gains explicitly, we introduce the effective gain vectors
$\boldsymbol{a}=\sqrt{\boldsymbol{\rho}}\odot\boldsymbol{\alpha}$
and
$\boldsymbol{b}=\sqrt{\mathbf{\text{1}}_{M}-\boldsymbol{\rho}}\odot\boldsymbol{\alpha}$.
The original parameters are recovered as
$\alpha_m=\sqrt{a_m^{\text{2}}+b_m^{\text{2}}}$
and
$\rho_m=a_m^{\text{2}}/(a_m^{\text{2}}+b_m^{\text{2}})$.
Accordingly, \eqref{eq:P0_gain} and \eqref{eq:P0_split} become
\begin{equation}
\text{1}\leq a_m^{\text{2}}+b_m^{\text{2}}
\leq\alpha_{\max}^{\text{2}},\quad
a_m,b_m\geq\text{0},\quad
\forall m\in\mathcal{M}.
\label{eq:branch_gain_constraint}
\end{equation}
Let
$\mathbf{B}_{\mathrm{c}}
=\mathrm{diag}(\boldsymbol{e}\odot\boldsymbol{\phi}_{\mathrm{a}}^{\mathrm{c}})$,
$\mathbf{B}_{\mathrm{s}}
=\mathrm{diag}(\boldsymbol{e}\odot\boldsymbol{\phi}_{\mathrm{a}}^{\mathrm{s}})$,
and
$\bar{\boldsymbol{\phi}}_{\mathrm{p}}
=(\mathbf{\text{1}}_{M}-\boldsymbol{e})\odot\boldsymbol{\phi}_{\mathrm{p}}$,
the active response matrices are then
$\boldsymbol{\Phi}_{\mathrm{a}}^{\mathrm{c}}
=\mathbf{B}_{\mathrm{c}}\mathrm{diag}(\boldsymbol{a})$
and
$\boldsymbol{\Phi}_{\mathrm{a}}^{\mathrm{s}}
=\mathbf{B}_{\mathrm{s}}\mathrm{diag}(\boldsymbol{b})$.
The full-aperture response vectors are
$\boldsymbol{\theta}_{\mathrm{c}}
=\mathbf{B}_{\mathrm{c}}\boldsymbol{a}+\bar{\boldsymbol{\phi}}_{\mathrm{p}}$
and
$\boldsymbol{\theta}_{\mathrm{s}}
=\mathbf{B}_{\mathrm{s}}\boldsymbol{b}+\bar{\boldsymbol{\phi}}_{\mathrm{p}}$.

For the expressions of power and rate, define
$\mathcal{Z}(\boldsymbol{h},\mathbf{R})
=\mathrm{diag}(\boldsymbol{h})\boldsymbol{G}\mathbf{R}
\boldsymbol{G}^{H}\mathrm{diag}(\boldsymbol{h}^{\ast})$
and
$\mathbf{D}_{p,q}^{\mathrm{c}}
=\mathbf{B}_{\mathrm{c}}^{H}
\mathrm{diag}(|\boldsymbol{h}_{p,q}|^{\text{2}})
\mathbf{B}_{\mathrm{c}}$.
To express the DL-UL power coupling in $P_{\mathrm{echo}}$, further define
$r_j=\|\boldsymbol{\Phi}_{\mathrm{a}}^{\mathrm{s}}
\boldsymbol{h}_{\mathrm{e},j}\|^{\text{2}}$,
$q_j=\|\boldsymbol{h}_{\mathrm{e},j}^{T}
\boldsymbol{\Theta}_{\mathrm{c}}\boldsymbol{G}
\mathbf{R}_{\mathrm{x}}^{\text{1}/\text{2}}\|^{\text{2}}$,
and
$t_j=\|\boldsymbol{h}_{\mathrm{e},j}^{T}
\boldsymbol{\Phi}_{\mathrm{a}}^{\mathrm{c}}\|^{\text{2}}$.
The amplified echo power is written as
\begin{equation}
P_{\mathrm{echo}}
=\sum_{j=\text{1}}^{J}|\varrho_{\mathrm{e},j}|^{\text{2}}
r_j(q_j+\delta_{\mathrm{a}}^{\text{2}}t_j).
\label{eq:branch_echo_output}
\end{equation}
With
$\mathbf{M}_{\mathrm{e},j}
=\mathcal{Z}(\boldsymbol{h}_{\mathrm{e},j},\mathbf{R}_{\mathrm{x}})$
and
$\mathbf{D}_{\mathrm{e},j}^{\mathrm{s}}
=\mathbf{B}_{\mathrm{s}}^{H}
\mathrm{diag}(|\boldsymbol{h}_{\mathrm{e},j}|^{\text{2}})
\mathbf{B}_{\mathrm{s}}$, these factors are expressed in the new variables $\boldsymbol{a}$ and $\boldsymbol{b}$ as follows
\begin{subequations}
\label{eq:r_t_j_1}
\begin{align}
&q_j(\boldsymbol{a})
=\boldsymbol{\theta}_{\mathrm{c}}^{T}
\mathbf{M}_{\mathrm{e},j}\boldsymbol{\theta}_{\mathrm{c}}^{\ast},\\
&t_j(\boldsymbol{a})
=\boldsymbol{a}^{T}\mathbf{D}_{\mathrm{e},j}^{\mathrm{c}}\boldsymbol{a},\\
&r_j(\boldsymbol{b})
=\boldsymbol{b}^{T}\mathbf{D}_{\mathrm{e},j}^{\mathrm{s}}\boldsymbol{b}.
\end{align}
\end{subequations}

Using \eqref{eq:branch_echo_output}, the UL branch power becomes
$P_{\mathrm{s}}(\boldsymbol{a},\boldsymbol{b})
=P_{\mathrm{echo}}(\boldsymbol{a},\boldsymbol{b})
+\delta_{\mathrm{a}}^{\text{2}}
\boldsymbol{b}^{T}\mathbf{D}_{\mathrm{s}}\boldsymbol{b}$,
where
$\mathbf{D}_{\mathrm{s}}=\mathbf{B}_{\mathrm{s}}^{H}\mathbf{B}_{\mathrm{s}}$.
The DL branch power is
$P_{\mathrm{c}}(\boldsymbol{a})
=\boldsymbol{a}^{T}\mathbf{D}_{\mathrm{c,x}}\boldsymbol{a}
+\delta_{\mathrm{a}}^{\text{2}}
\boldsymbol{a}^{T}\mathbf{D}_{\mathrm{c}}\boldsymbol{a}$,
where
$\mathbf{D}_{\mathrm{c,x}}
=\mathbf{B}_{\mathrm{c}}^{H}
\mathrm{diag}\!\left(
\mathrm{diag}(\boldsymbol{G}\mathbf{R}_{\mathrm{x}}\boldsymbol{G}^{H})
\right)\mathbf{B}_{\mathrm{c}}$
and
$\mathbf{D}_{\mathrm{c}}=\mathbf{B}_{\mathrm{c}}^{H}\mathbf{B}_{\mathrm{c}}$.
Therefore, the HRIS and total system powers are given by
\begin{subequations}
\label{eq:branch_total_power}
\begin{align}
P_{\mathrm{ris}}(\boldsymbol{a},\boldsymbol{b})
&=P_{\mathrm{sta}}+P_{\mathrm{c}}(\boldsymbol{a})
+P_{\mathrm{s}}(\boldsymbol{a},\boldsymbol{b}),
\label{eq:branch_hris_power_model}\\
P_{\mathrm{tot}}(\boldsymbol{a},\boldsymbol{b})
&=\mathrm{Tr}(\mathbf{R}_{\mathrm{x}})+P_{\mathrm{f}}
+P_{\mathrm{ris}}(\boldsymbol{a},\boldsymbol{b}),
\label{eq:branch_total_power_model}
\end{align}
\end{subequations}
where
$P_{\mathrm{sta}}\!=\!\left\Vert\boldsymbol{e}\right\Vert_{\text{1}}P_{\mathrm{a}}
\!+\!\left(M\!-\!\left\Vert\boldsymbol{e}\right\Vert_{\text{1}}\right)P_{\mathrm{p}}$
is the static circuit power.

Next, we express the rate constraints in terms of $\boldsymbol{a}$.
Define
$\mathbf{R}_{\mathrm{x},k}
=\boldsymbol{w}_{\mathrm{c},k}\boldsymbol{w}_{\mathrm{c},k}^{H}$,
$\mathbf{R}_{\mathrm{x},-k}
=\mathbf{R}_{\mathrm{x}}-\mathbf{R}_{\mathrm{x},k}$,
$\mathbf{Z}_{p,q,k}
=\mathcal{Z}(\boldsymbol{h}_{p,q},\mathbf{R}_{\mathrm{x},k})$,
and
$\mathbf{Z}_{p,q,-k}
=\mathcal{Z}(\boldsymbol{h}_{p,q},\mathbf{R}_{\mathrm{x},-k})$.
The signal and interference-plus-noise powers are
$D_{p,q,k}(\boldsymbol{a})
=\boldsymbol{\theta}_{\mathrm{c}}^{T}
\mathbf{Z}_{p,q,k}\boldsymbol{\theta}_{\mathrm{c}}^{\ast}$
and
$I_{p,q,k}(\boldsymbol{a})
=\boldsymbol{\theta}_{\mathrm{c}}^{T}
\mathbf{Z}_{p,q,-k}\boldsymbol{\theta}_{\mathrm{c}}^{\ast}
+\delta_{\mathrm{a}}^{\text{2}}
\boldsymbol{a}^{T}\mathbf{D}_{p,q}^{\mathrm{c}}\boldsymbol{a}
+\delta_{p,q}^{\text{2}}$,
respectively.
Both are convex quadratic functions of $\boldsymbol{a}$.
We therefore retain the user-interference term $I_{\mathrm{u},k}(\boldsymbol{a})$ and Eve-signal term $D_{\mathrm{e},j,k}(\boldsymbol{a})$ exactly, and construct affine lower bounds for the user-signal and Eve-interference terms.
For brevity, let $\mathbf{Z}_{\mathrm{u},k}
=\mathbf{Z}_{\mathrm{u},k,k}$ and
$\Delta\boldsymbol{a}=\boldsymbol{a}-\boldsymbol{a}^{(t)}$.
The lower bounds are given by
\begin{subequations}
\label{eq:gain_rate_lower_bounds}
\begin{align}
\underline{D}_{\mathrm{u},k}^{\mathrm{a},(t)}
={}&D_{\mathrm{u},k}(\boldsymbol{a}^{(t)})
+\text{2}\mathrm{Re}\left\{
\boldsymbol{\theta}_{\mathrm{c}}^{(t),T}
\mathbf{Z}_{\mathrm{u},k}
\mathbf{B}_{\mathrm{c}}^{\ast}\Delta\boldsymbol{a}
\right\},
\label{eq:a_user_signal_lb}\\
\underline{I}_{\mathrm{e},j,k}^{\mathrm{a},(t)}
={}&I_{\mathrm{e},j,k}(\boldsymbol{a}^{(t)})
+\text{2}\mathrm{Re}\left\{
\boldsymbol{\theta}_{\mathrm{c}}^{(t),T}
\mathbf{Z}_{\mathrm{e},j,-k}
\mathbf{B}_{\mathrm{c}}^{\ast}\Delta\boldsymbol{a}
\right\}
\nonumber\\
&+\text{2}\delta_{\mathrm{a}}^{\text{2}}
\boldsymbol{a}^{(t),T}
\mathbf{D}_{\mathrm{e},j}^{\mathrm{c}}\Delta\boldsymbol{a}.
\label{eq:a_eve_interference_lb}
\end{align}
\end{subequations}
With these expressions and bounds,
\eqref{eq:interference_split} and \eqref{eq:signal_split1}
are conservatively approximated by
\begin{subequations}
\label{eq:beam_sinr1}
\begin{alignat}{2}
&\underline{D}_{\mathrm{u},k}^{\mathrm{a},(t)}
\geq g_{\mathrm{u},k}^{\mathrm{ub},(t)},\ 
&&D_{\mathrm{e},j,k}(\boldsymbol{a})
\leq g_{\mathrm{e},j,k}^{\mathrm{lb},(t)},
\forall j\in\mathcal{J},\forall k\in\mathcal{K},
\label{eq:a_signal_constraints}\\
&I_{\mathrm{u},k}(\boldsymbol{a})\leq\varpi_{\mathrm{u},k},\ 
&&\underline{I}_{\mathrm{e},j,k}^{\mathrm{a},(t)}
\geq\varpi_{\mathrm{e},j,k},
\forall j\in\mathcal{J},\forall k\in\mathcal{K}.
\label{eq:a_interference_constraints}
\end{alignat}
\end{subequations}

For the CRB constraint \eqref{eq:crb_schur}, define
$\mathbf{U}(\boldsymbol{a},\boldsymbol{b})
=\left[
\mathrm{vec}(\mathbf{H}'_{\nu}
\mathbf{R}_{\mathrm{x}}^{\text{1}/\text{2}})
\right]_{\nu=\text{1}}^{\text{5}J}$
and
$\mathbf{Y}(\boldsymbol{b})
=\mathbf{I}_{N}\otimes\mathbf{R}_{\omega}(\boldsymbol{b})$.
Here, $\mathbf{H}'_{\nu}$ denotes the channel derivative associated with the $\nu$-th parameter in $\boldsymbol{\zeta}$, and $\mathbf{R}_{\omega}\left(\boldsymbol{b}\right)
=\delta_{\mathrm{a}}^{\text{2}}
\bar{\boldsymbol{G}}\mathbf{B}_{\mathrm{s}}
\mathrm{diag}\left(\boldsymbol{b}\odot\boldsymbol{b}\right)
\mathbf{B}_{\mathrm{s}}^{H}\bar{\boldsymbol{G}}^{H}
+\mathbf{R}_{\mathrm{BR}}$.
The FIM is then
$\mathbf{F}=\text{2}L\mathrm{Re}
\{\mathbf{U}^{H}\mathbf{Y}^{-\text{1}}\mathbf{U}\}$.
For each branch update, let
$\mathbf{K}^{(t)}
=(\mathbf{Y}^{(t)})^{-\text{1}}\mathbf{U}^{(t)}$,
where $\mathbf{U}^{(t)}$ and $\mathbf{Y}^{(t)}$
are evaluated at the current design.
An affine matrix lower bound, tight at the reference point
$\left(\mathbf{U}^{(t)},\mathbf{Y}^{(t)}\right)$, is given by
\begin{equation}
\underline{\mathbf{F}}^{(t)}\left(\mathbf{U},\mathbf{Y}\right)
=\text{2}L\mathrm{Re}\Bigl\{
\mathbf{K}^{(t),H}\mathbf{U}
+\mathbf{U}^{H}\mathbf{K}^{(t)}
-\mathbf{K}^{(t),H}\mathbf{Y}\mathbf{K}^{(t)}
\Bigr\}.
\label{eq:branch_fim_lower}
\end{equation}
We update $\boldsymbol{a}$ and $\boldsymbol{b}$ successively to handle their coupling in the power expressions, FIM, and gain constraint \eqref{eq:branch_gain_constraint}.

\subsubsection{DL Branch Gain Optimization}

With $\boldsymbol{b}$ fixed, the rate constraints are handled by \eqref{eq:beam_sinr1}.
Since $r_j(\boldsymbol{b})$ is constant, the HRIS and total powers in \eqref{eq:branch_total_power} are convex quadratic functions of $\boldsymbol{a}$.
We denote these fixed-$\boldsymbol{b}$ expressions by
$P_{\mathrm{ris}}(\boldsymbol{a})$ and
$P_{\mathrm{tot}}(\boldsymbol{a})$, respectively.
For sensing, $\mathbf{U}(\boldsymbol{a})$ is affine and
$\mathbf{Y}=\mathbf{Y}^{(t)}$ is fixed. 
Substituting them into \eqref{eq:branch_fim_lower} gives the affine FIM lower bound
$\underline{\mathbf{F}}_{\mathrm{a}}^{(t)}$. The gain constraint \eqref{eq:branch_gain_constraint} also reduces to an interval for each $a_m$.
The convex DL gain subproblem is given by
\begin{subequations}
\label{eq:P7}
\begin{eqnarray}
\mathrm{P\text{7}}:
&\underset{\boldsymbol{a},\boldsymbol{\Delta}}{\max}
&\psi
\label{eq:P7_obj}\\
&\mathrm{s.t.}&
\eqref{eq:P2_qos}-\eqref{eq:P2_positive},
\eqref{eq:crb_trace_aux},\eqref{eq:beam_sinr1},
\label{eq:P7_1}\\
&&
\!\!\sqrt{\left[\text{1}\!\!-\!\!b_m^{\text{2}}\right]^+}\!\leq \!a_m\!\!\leq\!\!\sqrt{\alpha_{\max}^{\text{2}}\!\!-\!\!b_m^{\text{2}}},\!\forall m\!\in\!\mathcal{M},
\label{eq:a_gain_interval}\\
&&
P_{\mathrm{ris}}(\boldsymbol{a})\leq P_{\mathrm{ris},\max},
\ 
P_{\mathrm{tot}}(\boldsymbol{a})\leq\eta,
\label{eq:a_hris_power}\\
&&
\begin{bmatrix}
\underline{\mathbf{F}}_{\mathrm{a}}^{(t)}&\mathbf{J}_{j}^{T}\\
\mathbf{J}_{j}&\mathbf{C}_{j}
\end{bmatrix}
\succeq\boldsymbol{\text{0}},
\ \forall j\in\mathcal{J}.
\label{eq:a_crb_lmi}
\end{eqnarray}
\end{subequations}

\subsubsection{UL Branch Gain Optimization}

With the updated $\boldsymbol{a}$ fixed, the communication powers are independent of $\boldsymbol{b}$.
We therefore retain the feasible rate bounds
$v_{\mathrm{u},k}^{(t)}$ and $v_{\mathrm{e},k}^{(t)}$
from the preceding DL update in \eqref{eq:P2_see}, which continues to couple $\psi$ and $\eta$.
Moreover, $P_{\mathrm{c}}$, $q_j$, and $t_j$ are fixed, whereas $r_j(\boldsymbol{b})$ is quadratic.
Thus, the fixed-$\boldsymbol{a}$ power expressions in \eqref{eq:branch_total_power}, denoted by
$P_{\mathrm{ris}}(\boldsymbol{b})$ and
$P_{\mathrm{tot}}(\boldsymbol{b})$, are convex quadratic functions of $\boldsymbol{b}$.

For sensing, $\mathbf{U}(\boldsymbol{b})$ is affine in
$\boldsymbol{b}$, whereas $\mathbf{Y}(\boldsymbol{b})$
contains the quadratic terms $b_m^{\text{2}}$.
Therefore, direct substitution into \eqref{eq:branch_fim_lower}  gives a non-affine FIM lower bound.
We introduce $\boldsymbol{z}=[z_{\text{1}},\dots,z_M]^T$
with $z_m\geq b_m^{\text{2}}$ to replace these quadratic
terms with auxiliary upper bounds.
This gives the affine covariance upper bound
$\overline{\mathbf{R}}_{\omega}(\boldsymbol{z})
=\delta_{\mathrm{a}}^{\text{2}}
\bar{\boldsymbol{G}}\mathbf{B}_{\mathrm{s}}
\mathrm{diag}(\boldsymbol{z})
\mathbf{B}_{\mathrm{s}}^{H}\bar{\boldsymbol{G}}^{H}
+\mathbf{R}_{\mathrm{BR}}$
and
$\overline{\mathbf{Y}}(\boldsymbol{z})
=\mathbf{I}_{N}\otimes
\overline{\mathbf{R}}_{\omega}(\boldsymbol{z})$.
Since
$\overline{\mathbf{Y}}(\boldsymbol{z})
\succeq\mathbf{Y}(\boldsymbol{b})$,
applying \eqref{eq:branch_fim_lower} to
$\mathbf{U}(\boldsymbol{b})$ and
$\overline{\mathbf{Y}}(\boldsymbol{z})$
gives a conservative affine FIM lower bound
$\underline{\mathbf{F}}_{\mathrm{b}}^{(t)}$.
The reference matrices are evaluated using the updated $\boldsymbol{a}$ and current $\boldsymbol{b}^{(t)}$, with
$z_m^{(t)}=(b_m^{(t)})^{\text{2}}$.
The convex UL gain subproblem is given by
\begin{subequations}
\label{eq:P8}
\begin{eqnarray}
\mathrm{P\text{8}}:
&\underset{\substack{
\boldsymbol{b},\boldsymbol{z},\psi,\\
\eta,\{\mathbf{C}_{j}\}
}}{\max}
&\psi
\label{eq:P8_obj}\\
&\mathrm{s.t.}&
\sum_{k=\text{1}}^{K}
\left(v_{\mathrm{u},k}^{(t)}-v_{\mathrm{e},k}^{(t)}\right)
\geq\mathcal{G}^{(t)}(\psi,\eta),
\label{eq:b_fixed_rate_see}\\
&&
\!\!\!\!\sqrt{\!\left[\text{1}\!-\!a_m^{\text{2}}\right]^+}\!\!\leq \!b_m\!\leq\!\!\sqrt{\alpha_{\max}^{\text{2}}\!-\!a_m^{\text{2}}},\!\forall m\!\in\!\mathcal{M},
\label{eq:b_gain_interval}\\
&&
P_{\mathrm{ris}}(\boldsymbol{b})\leq P_{\mathrm{ris},\max},
\ 
P_{\mathrm{tot}}(\boldsymbol{b})\leq\eta,
\label{eq:b_total_power}\\
&&
\begin{bmatrix}
z_m&b_m\\
b_m&\text{1}
\end{bmatrix}
\succeq\boldsymbol{\text{0}},
\ \forall m\in\mathcal{M},
\label{eq:P8_z}\\
&&
\begin{bmatrix}
\underline{\mathbf{F}}_{\mathrm{b}}^{(t)}&\mathbf{J}_{j}^{T}\\
\mathbf{J}_{j}&\mathbf{C}_{j}
\end{bmatrix}
\succeq\boldsymbol{\text{0}},
\ \forall j\in\mathcal{J},
\label{eq:b_crb_lmi}\\
&&
\eqref{eq:crb_trace_aux}.
\label{eq:P8_trace}
\end{eqnarray}
\end{subequations}
After solving $\mathrm{P\text{7}}$ and $\mathrm{P\text{8}}$ using CVX, we recover
$\boldsymbol{\alpha}$ and $\boldsymbol{\rho}$
from the updated $\boldsymbol{a}$ and $\boldsymbol{b}$
using the relations above.

%%%%%%%%%%%%%%%%%%%%%%%%%%%%%%%%%%%%%%%%%%%%
%%%%%%%%%%%%%%%%%%%%%%%%%%%%%%%%%%%%%%%%%%%%
%%%%%%%%%%%%%%%%%%%%%%%%%%%%%%%%%%%%%%%%%%%%
%%%%%%%%%%%%%%%%%%%%%%%%%%%%%%%%%%%%%%%%%%%%\vspace{-4mm}
\subsection{Active Phase-Shift Optimization}

With the other design variables fixed, the DL and UL active phase vectors are updated by solving
\begin{subequations}
\label{eq:P9}
\begin{eqnarray}
\mathrm{P\text{9}}:
&\underset{\boldsymbol{\phi}_{\mathrm{a}}^{\mathrm{c}},
\boldsymbol{\phi}_{\mathrm{a}}^{\mathrm{s}},
\boldsymbol{\Delta}}{\max}
&\psi
\label{eq:P9_obj}\\
&\mathrm{s.t.}&
\eqref{eq:P0_hris_power},\eqref{eq:P0_active_phase},
\eqref{eq:P1_power},\eqref{eq:P2_qos}-\eqref{eq:P2_positive},
\label{eq:P9_1}\\
&&
\eqref{eq:interference_split},\eqref{eq:signal_split1},
\eqref{eq:crb_schur}.
\label{eq:P9_2}
\end{eqnarray}
\end{subequations}

We update the two phase vectors successively in vector form using PCCP~\cite{r_PCCP}.
For either phase vector
$\boldsymbol{\phi}\in
\{\boldsymbol{\phi}_{\mathrm{a}}^{\mathrm{c}},
\boldsymbol{\phi}_{\mathrm{a}}^{\mathrm{s}}\}$,
we relax the unit-modulus constraints \eqref{eq:P0_active_phase} to
$|\phi_m|^{\text{2}}\leq\text{1}$
and penalize the residual
$\mathcal{R}_{\phi}(\boldsymbol{\phi})
=\sum_{m=\text{1}}^{M}(\text{1}-|\phi_m|^{\text{2}})$
to promote unit-modulus solutions.
At the reference point $\boldsymbol{\phi}^{(t)}$,
a tight affine upper bound on this residual is
\begin{equation}
\overline{\mathcal{R}}_{\phi}^{(t)}(\boldsymbol{\phi})
=\sum_{m=\text{1}}^{M}
\left[
\text{1}
-\text{2}\mathrm{Re}\{\phi_m^{(t),\ast}\phi_m\}
+|\phi_m^{(t)}|^{\text{2}}
\right].
\label{eq:phase_penalty}
\end{equation}
To express the HRIS responses in terms of the phase vectors, let
$\mathbf{B}_{\mathrm{c}}^{\phi}
=\mathrm{diag}(\boldsymbol{e}\odot\boldsymbol{a})$
and
$\mathbf{B}_{\mathrm{s}}^{\phi}
=\mathrm{diag}(\boldsymbol{e}\odot\boldsymbol{b})$,
which are fixed in this subproblem.
The full-aperture response vectors are then
$\boldsymbol{\theta}_{\mathrm{c}}
=\mathbf{B}_{\mathrm{c}}^{\phi}
\boldsymbol{\phi}_{\mathrm{a}}^{\mathrm{c}}
+\bar{\boldsymbol{\phi}}_{\mathrm{p}}$
and
$\boldsymbol{\theta}_{\mathrm{s}}
=\mathbf{B}_{\mathrm{s}}^{\phi}
\boldsymbol{\phi}_{\mathrm{a}}^{\mathrm{s}}
+\bar{\boldsymbol{\phi}}_{\mathrm{p}}$.

For the expressions of power and rate, define
$\mathbf{D}_{p,q}^{\mathrm{c},\phi}
=(\mathbf{B}_{\mathrm{c}}^{\phi})^{H}
\mathrm{diag}(|\boldsymbol{h}_{p,q}|^{\text{2}})
\mathbf{B}_{\mathrm{c}}^{\phi}$
and
$\mathbf{D}_{\mathrm{e},j}^{\mathrm{s},\phi}
=(\mathbf{B}_{\mathrm{s}}^{\phi})^{H}
\mathrm{diag}(|\boldsymbol{h}_{\mathrm{e},j}|^{\text{2}})
\mathbf{B}_{\mathrm{s}}^{\phi}$.
The matrices
$\mathbf{D}_{\mathrm{c,x}}^{\phi}$,
$\mathbf{D}_{\mathrm{c}}^{\phi}$, and
$\mathbf{D}_{\mathrm{s}}^{\phi}$
follow their definitions in the preceding subsection, with
$\mathbf{B}_{\mathrm{c}}$ and $\mathbf{B}_{\mathrm{s}}$
replaced by
$\mathbf{B}_{\mathrm{c}}^{\phi}$ and
$\mathbf{B}_{\mathrm{s}}^{\phi}$, respectively.
The power factors in \eqref{eq:r_t_j_1} become
$q_j=\boldsymbol{\theta}_{\mathrm{c}}^{T}
\mathbf{M}_{\mathrm{e},j}\boldsymbol{\theta}_{\mathrm{c}}^{\ast}$,
$t_j=(\boldsymbol{\phi}_{\mathrm{a}}^{\mathrm{c}})^{H}
\mathbf{D}_{\mathrm{e},j}^{\mathrm{c},\phi}
\boldsymbol{\phi}_{\mathrm{a}}^{\mathrm{c}}$,
and
$r_j=(\boldsymbol{\phi}_{\mathrm{a}}^{\mathrm{s}})^{H}
\mathbf{D}_{\mathrm{e},j}^{\mathrm{s},\phi}
\boldsymbol{\phi}_{\mathrm{a}}^{\mathrm{s}}$.
Using these factors, the DL and UL branch powers are
\begin{subequations}
\label{eq:active_phase_branch_powers}
\begin{align}
P_{\mathrm{c}}(\boldsymbol{\phi}_{\mathrm{a}}^{\mathrm{c}})
=&
(\boldsymbol{\phi}_{\mathrm{a}}^{\mathrm{c}})^{H}
\left(
\mathbf{D}_{\mathrm{c,x}}^{\phi}
+\delta_{\mathrm{a}}^{\text{2}}\mathbf{D}_{\mathrm{c}}^{\phi}
\right)
\boldsymbol{\phi}_{\mathrm{a}}^{\mathrm{c}},\\
P_{\mathrm{s}}(
\boldsymbol{\phi}_{\mathrm{a}}^{\mathrm{c}},
\boldsymbol{\phi}_{\mathrm{a}}^{\mathrm{s}})
=&
\sum_{j=\text{1}}^{J}|\varrho_{\mathrm{e},j}|^{\text{2}}
r_j(q_j+\delta_{\mathrm{a}}^{\text{2}}t_j)+\delta_{\mathrm{a}}^{\text{2}}
(\boldsymbol{\phi}_{\mathrm{a}}^{\mathrm{s}})^{H}
\mathbf{D}_{\mathrm{s}}^{\phi}
\boldsymbol{\phi}_{\mathrm{a}}^{\mathrm{s}}.
\end{align}
\end{subequations}
The corresponding HRIS and total powers are
$P_{\mathrm{ris}}(
\boldsymbol{\phi}_{\mathrm{a}}^{\mathrm{c}},
\boldsymbol{\phi}_{\mathrm{a}}^{\mathrm{s}})
=P_{\mathrm{sta}}
+P_{\mathrm{c}}(\boldsymbol{\phi}_{\mathrm{a}}^{\mathrm{c}})
+P_{\mathrm{s}}(
\boldsymbol{\phi}_{\mathrm{a}}^{\mathrm{c}},
\boldsymbol{\phi}_{\mathrm{a}}^{\mathrm{s}})$
and
$P_{\mathrm{tot}}(
\boldsymbol{\phi}_{\mathrm{a}}^{\mathrm{c}},
\boldsymbol{\phi}_{\mathrm{a}}^{\mathrm{s}})
=\mathrm{Tr}(\mathbf{R}_{\mathrm{x}})+P_{\mathrm{f}}
+P_{\mathrm{ris}}(
\boldsymbol{\phi}_{\mathrm{a}}^{\mathrm{c}},
\boldsymbol{\phi}_{\mathrm{a}}^{\mathrm{s}})$.

Next, we address the rate constraints
\eqref{eq:interference_split} and \eqref{eq:signal_split1}.
Using the matrices $\mathbf{Z}_{p,q,k}$ and
$\mathbf{Z}_{p,q,-k}$ defined in the preceding subsection,
the signal and interference-plus-noise powers are
\begin{subequations}
\label{eq:active_phase_received_powers}
\begin{align}
D_{p,q,k}(\boldsymbol{\phi}_{\mathrm{a}}^{\mathrm{c}})
=&
\boldsymbol{\theta}_{\mathrm{c}}^{T}
\mathbf{Z}_{p,q,k}\boldsymbol{\theta}_{\mathrm{c}}^{\ast},
\label{eq:active_phase_signal_power}\\
I_{p,q,k}(\boldsymbol{\phi}_{\mathrm{a}}^{\mathrm{c}})
=&
\boldsymbol{\theta}_{\mathrm{c}}^{T}
\mathbf{Z}_{p,q,-k}\boldsymbol{\theta}_{\mathrm{c}}^{\ast}+\delta_{\mathrm{a}}^{\text{2}}
(\boldsymbol{\phi}_{\mathrm{a}}^{\mathrm{c}})^{H}
\mathbf{D}_{p,q}^{\mathrm{c},\phi}
\boldsymbol{\phi}_{\mathrm{a}}^{\mathrm{c}}
+\delta_{p,q}^{\text{2}}.
\label{eq:active_phase_interference_power}
\end{align}
\end{subequations}
Both expressions are convex quadratic functions of
$\boldsymbol{\phi}_{\mathrm{a}}^{\mathrm{c}}$.
With
$\Delta\boldsymbol{\phi}_{\mathrm{c}}
=\boldsymbol{\phi}_{\mathrm{a}}^{\mathrm{c}}
-\boldsymbol{\phi}_{\mathrm{a}}^{\mathrm{c},(t)}$, the lower bounds of user-signal and
Eve-interference terms are
\begin{subequations}
\label{eq:active_phase_rate_lower_bounds}
\begin{align}
\underline{D}_{\mathrm{u},k}^{\mathrm{c},(t)}
={}&
D_{\mathrm{u},k}
(\boldsymbol{\phi}_{\mathrm{a}}^{\mathrm{c},(t)})
+\text{2}\mathrm{Re}\left\{
\boldsymbol{\theta}_{\mathrm{c}}^{(t),T}
\mathbf{Z}_{\mathrm{u},k}
(\mathbf{B}_{\mathrm{c}}^{\phi})^{\ast}
\Delta\boldsymbol{\phi}_{\mathrm{c}}^{\ast}
\right\},
\label{eq:active_dl_user_lb}\\
\underline{I}_{\mathrm{e},j,k}^{\mathrm{c},(t)}
={}&
I_{\mathrm{e},j,k}
(\boldsymbol{\phi}_{\mathrm{a}}^{\mathrm{c},(t)})
+\text{2}\mathrm{Re}\left\{
\boldsymbol{\theta}_{\mathrm{c}}^{(t),T}
\mathbf{Z}_{\mathrm{e},j,-k}
(\mathbf{B}_{\mathrm{c}}^{\phi})^{\ast}
\Delta\boldsymbol{\phi}_{\mathrm{c}}^{\ast}
\right\}
\nonumber\\
&+\text{2}\delta_{\mathrm{a}}^{\text{2}}
\mathrm{Re}\left\{
(\boldsymbol{\phi}_{\mathrm{a}}^{\mathrm{c},(t)})^{H}
\mathbf{D}_{\mathrm{e},j}^{\mathrm{c},\phi}
\Delta\boldsymbol{\phi}_{\mathrm{c}}
\right\}.
\label{eq:active_dl_eve_lb}
\end{align}
\end{subequations}
With these expressions and bounds, the rate constraints are
conservatively approximated by
\begin{subequations}
\label{eq:active_dl_comm}
\begin{alignat}{2}
&\underline{D}_{\mathrm{u},k}^{\mathrm{c},(t)}
\geq g_{\mathrm{u},k}^{\mathrm{ub},(t)},
&&D_{\mathrm{e},j,k}
(\boldsymbol{\phi}_{\mathrm{a}}^{\mathrm{c}})
\leq g_{\mathrm{e},j,k}^{\mathrm{lb},(t)},
\forall j\in\mathcal{J},\forall k\in\mathcal{K},
\label{eq:active_dl_signal_constraints}\\
&I_{\mathrm{u},k}
(\boldsymbol{\phi}_{\mathrm{a}}^{\mathrm{c}})
\leq\varpi_{\mathrm{u},k},\ 
&&\underline{I}_{\mathrm{e},j,k}^{\mathrm{c},(t)}
\geq\varpi_{\mathrm{e},j,k},
\forall j\in\mathcal{J},\forall k\in\mathcal{K}.
\label{eq:active_dl_interference_constraints}
\end{alignat}
\end{subequations}

For sensing, the disturbance covariance is
$\mathbf{R}_{\omega}
(\boldsymbol{\phi}_{\mathrm{a}}^{\mathrm{s}})
=\delta_{\mathrm{a}}^{\text{2}}
\bar{\boldsymbol{G}}\mathbf{B}_{\mathrm{s}}^{\phi}
\mathrm{diag}(
|\boldsymbol{\phi}_{\mathrm{a}}^{\mathrm{s}}|^{\text{2}})
(\mathbf{B}_{\mathrm{s}}^{\phi})^{H}
\bar{\boldsymbol{G}}^{H}
+\mathbf{R}_{\mathrm{BR}}$,
with
$\mathbf{Y}(\boldsymbol{\phi}_{\mathrm{a}}^{\mathrm{s}})
=\mathbf{I}_{N}\otimes
\mathbf{R}_{\omega}(\boldsymbol{\phi}_{\mathrm{a}}^{\mathrm{s}})$.
We use \eqref{eq:branch_fim_lower} to construct the FIM lower bounds,
with the reference matrices evaluated at the current design.

\subsubsection{DL Active Phase Optimization}

With $\boldsymbol{\phi}_{\mathrm{a}}^{\mathrm{s}}$ fixed,
$r_j$ is constant, so the HRIS and total powers are convex
quadratic functions of $\boldsymbol{\phi}_{\mathrm{a}}^{\mathrm{c}}$.
We denote these fixed-$\boldsymbol{\phi}_{\mathrm{a}}^{\mathrm{s}}$ expressions by
$P_{\mathrm{ris}}(\boldsymbol{\phi}_{\mathrm{a}}^{\mathrm{c}})$
and
$P_{\mathrm{tot}}(\boldsymbol{\phi}_{\mathrm{a}}^{\mathrm{c}})$,
respectively.
The rate constraints are handled by \eqref{eq:active_dl_comm}.
For sensing, $\boldsymbol{\Theta}_{\mathrm{s}}$ and
$\mathbf{R}_{\omega}$ are fixed, so
$\mathbf{U}(\boldsymbol{\phi}_{\mathrm{a}}^{\mathrm{c}})$
is affine.
Applying \eqref{eq:branch_fim_lower} with this matrix and
$\mathbf{Y}=\mathbf{Y}^{(t)}$ gives the affine FIM lower bound
$\underline{\mathbf{F}}_{\phi_{\mathrm{c}}}^{(t)}$.
The convex PCCP subproblem for
$\boldsymbol{\phi}_{\mathrm{a}}^{\mathrm{c}}$ is given by
\begin{subequations}
\label{eq:P10}
\begin{eqnarray}
\mathrm{P\text{10}}:
&\underset{\boldsymbol{\phi}_{\mathrm{a}}^{\mathrm{c}},
\boldsymbol{\Delta}}{\max}
&
\psi-\tau_{\mathrm{c}}
\overline{\mathcal{R}}_{\phi}^{(t)}
(\boldsymbol{\phi}_{\mathrm{a}}^{\mathrm{c}})
\label{eq:P10_obj}\\
&\mathrm{s.t.}&
\eqref{eq:P2_qos}-\eqref{eq:P2_positive},
\eqref{eq:crb_trace_aux},\eqref{eq:active_dl_comm},
\label{eq:P10_1}\\
&&
P_{\mathrm{ris}}
(\boldsymbol{\phi}_{\mathrm{a}}^{\mathrm{c}})
\leq P_{\mathrm{ris},\max},
\label{eq:P10_hris_power}\\
&&
P_{\mathrm{tot}}
(\boldsymbol{\phi}_{\mathrm{a}}^{\mathrm{c}})
\leq\eta,
\label{eq:P10_total_power}\\
&&
\begin{bmatrix}
\underline{\mathbf{F}}_{\phi_{\mathrm{c}}}^{(t)}
&\mathbf{J}_j^{T}\\
\mathbf{J}_j&\mathbf{C}_j
\end{bmatrix}
\succeq\boldsymbol{\text{0}},
\ \forall j\in\mathcal{J},
\label{eq:P10_crb}\\
&&
|[\boldsymbol{\phi}_{\mathrm{a}}^{\mathrm{c}}]_m|^{\text{2}}
\leq\text{1},
\ \forall m\in\mathcal{M}.
\label{eq:P10_phase}
\end{eqnarray}
\end{subequations}
Here, $\tau_{\mathrm{c}}>\text{0}$ is the DL phase penalty weight.

\subsubsection{UL Active Phase Optimization}

With the updated $\boldsymbol{\phi}_{\mathrm{a}}^{\mathrm{c}}$
fixed, $P_{\mathrm{c}}$, $q_j$, and $t_j$ are constant,
whereas $r_j$ is quadratic in
$\boldsymbol{\phi}_{\mathrm{a}}^{\mathrm{s}}$.
The corresponding HRIS and total powers, denoted by
$P_{\mathrm{ris}}(\boldsymbol{\phi}_{\mathrm{a}}^{\mathrm{s}})$
and
$P_{\mathrm{tot}}(\boldsymbol{\phi}_{\mathrm{a}}^{\mathrm{s}})$,
are therefore convex quadratic functions.
The communication rates are fixed in this update.
Thus, \eqref{eq:P2_user_rate} and \eqref{eq:P2_eve_rate}
are retained as affine constraints in the rate bounds
$\boldsymbol{v}$.

Under the relaxed unit-modulus constraints, the squared
magnitudes in
$\mathbf{R}_{\omega}(\boldsymbol{\phi}_{\mathrm{a}}^{\mathrm{s}})$
remain variable.
To construct an affine covariance upper bound, we introduce
$\boldsymbol{z}_{\mathrm{s}}
=[z_{\mathrm{s},\text{1}},\dots,z_{\mathrm{s},M}]^T$
and replace each squared magnitude
$|[\boldsymbol{\phi}_{\mathrm{a}}^{\mathrm{s}}]_m|^{\text{2}}$
with an auxiliary upper bound
$z_{\mathrm{s},m}
\geq|[\boldsymbol{\phi}_{\mathrm{a}}^{\mathrm{s}}]_m|^{\text{2}}$.
This gives
$\overline{\mathbf{R}}_{\omega}(\boldsymbol{z}_{\mathrm{s}})
=\delta_{\mathrm{a}}^{\text{2}}
\bar{\boldsymbol{G}}\mathbf{B}_{\mathrm{s}}^{\phi}
\mathrm{diag}(\boldsymbol{z}_{\mathrm{s}})
(\mathbf{B}_{\mathrm{s}}^{\phi})^{H}
\bar{\boldsymbol{G}}^{H}
+\mathbf{R}_{\mathrm{BR}}$
and
$\overline{\mathbf{Y}}(\boldsymbol{z}_{\mathrm{s}})
=\mathbf{I}_{N}\otimes
\overline{\mathbf{R}}_{\omega}(\boldsymbol{z}_{\mathrm{s}})$.
Since
$\mathbf{U}(\boldsymbol{\phi}_{\mathrm{a}}^{\mathrm{s}})$
is affine and
$\overline{\mathbf{Y}}(\boldsymbol{z}_{\mathrm{s}})
\succeq\mathbf{Y}(\boldsymbol{\phi}_{\mathrm{a}}^{\mathrm{s}})$,
applying \eqref{eq:branch_fim_lower} gives the conservative
affine FIM lower bound
$\underline{\mathbf{F}}_{\phi_{\mathrm{s}}}^{(t)}$.
It is tight at the current reference point when
$z_{\mathrm{s},m}^{(t)}
=|[\boldsymbol{\phi}_{\mathrm{a}}^{\mathrm{s},(t)}]_m|^{\text{2}}$.
Let
$\boldsymbol{\Delta}_{\mathrm{s}}
=\{\psi,\eta,\boldsymbol{v},\{\mathbf{C}_j\}\}$.
The convex PCCP subproblem for
$\boldsymbol{\phi}_{\mathrm{a}}^{\mathrm{s}}$ is given by
\begin{subequations}
\label{eq:P11}
\begin{eqnarray}
\mathrm{P\text{11}}:
&\underset{\boldsymbol{\phi}_{\mathrm{a}}^{\mathrm{s}},
\boldsymbol{z}_{\mathrm{s}},
\boldsymbol{\Delta}_{\mathrm{s}}}{\max}
&
\psi-\tau_{\mathrm{s}}
\overline{\mathcal{R}}_{\phi}^{(t)}
(\boldsymbol{\phi}_{\mathrm{a}}^{\mathrm{s}})
\label{eq:P11_obj}\\
&\mathrm{s.t.}&
\eqref{eq:P2_user_rate}-\eqref{eq:P2_positive},
\eqref{eq:crb_trace_aux},
\label{eq:P11_rate}\\
&&
P_{\mathrm{ris}}
(\boldsymbol{\phi}_{\mathrm{a}}^{\mathrm{s}})
\leq P_{\mathrm{ris},\max},
\label{eq:P11_hris_power}\\
&&
P_{\mathrm{tot}}
(\boldsymbol{\phi}_{\mathrm{a}}^{\mathrm{s}})
\leq\eta,
\label{eq:P11_total_power}\\
&&
\begin{bmatrix}
\underline{\mathbf{F}}_{\phi_{\mathrm{s}}}^{(t)}
&\mathbf{J}_j^{T}\\
\mathbf{J}_j&\mathbf{C}_j
\end{bmatrix}
\succeq\boldsymbol{\text{0}},
\ \forall j\in\mathcal{J},
\label{eq:P11_crb}\\
&&
|[\boldsymbol{\phi}_{\mathrm{a}}^{\mathrm{s}}]_m|^{\text{2}}
\leq\text{1},
\ \forall m\in\mathcal{M},
\label{eq:P11_phase}\\
&&
\begin{bmatrix}
z_{\mathrm{s},m}
&[\boldsymbol{\phi}_{\mathrm{a}}^{\mathrm{s}}]_m\\
[\boldsymbol{\phi}_{\mathrm{a}}^{\mathrm{s}}]_m^{\ast}
&\text{1}
\end{bmatrix}
\succeq\boldsymbol{\text{0}},
\ \forall m\in\mathcal{M}.
\label{eq:P11_z}
\end{eqnarray}
\end{subequations}
Here, $\tau_{\mathrm{s}}>\text{0}$ is the UL phase penalty weight.
Both phase subproblems use inner PCCP iterations with updated
reference phases and adaptive penalty weights.

%%%%%%%%%%%%%%%%%%%%%%%%%%%%%%%%%%%%%%%%%%%%\vspace{-2mm}
\subsection{Passive Phase-Shift Optimization}

With the other design variables fixed, the passive phase vector is updated by solving
\begin{subequations}
\label{eq:P12}
\begin{eqnarray}
\mathrm{P\text{12}}:
&\underset{\boldsymbol{\phi}_{\mathrm{p}},\boldsymbol{\Delta}}{\max}
&\psi
\label{eq:P12_obj}\\
&\mathrm{s.t.}&
\eqref{eq:P0_hris_power},\eqref{eq:P0_passive_phase},
\eqref{eq:P1_power},\eqref{eq:P2_qos}-\eqref{eq:P2_positive},
\label{eq:P12_1}\\
&&
\eqref{eq:interference_split},\eqref{eq:signal_split1},
\eqref{eq:crb_schur}.
\label{eq:P12_3}
\end{eqnarray}
\end{subequations}
Let
$\mathbf{B}_{\mathrm{p}}
=\mathrm{diag}(\mathbf{\text{1}}_{M}-\boldsymbol{e})$.
The full-aperture HRIS responses are then
$\boldsymbol{\theta}_{\mathrm{c}}
=\mathbf{B}_{\mathrm{c}}^{\phi}
\boldsymbol{\phi}_{\mathrm{a}}^{\mathrm{c}}
+\mathbf{B}_{\mathrm{p}}\boldsymbol{\phi}_{\mathrm{p}}$
and
$\boldsymbol{\theta}_{\mathrm{s}}
=\mathbf{B}_{\mathrm{s}}^{\phi}
\boldsymbol{\phi}_{\mathrm{a}}^{\mathrm{s}}
+\mathbf{B}_{\mathrm{p}}\boldsymbol{\phi}_{\mathrm{p}}$.
We relax \eqref{eq:P0_passive_phase} to
$|[\boldsymbol{\phi}_{\mathrm{p}}]_m|^{\text{2}}\leq\text{1}$
and use the affine penalty bound \eqref{eq:phase_penalty}
with $\boldsymbol{\phi}=\boldsymbol{\phi}_{\mathrm{p}}$.
For the power constraints \eqref{eq:P0_hris_power} and
\eqref{eq:P1_power}, $P_{\mathrm{c}}$, $r_j$, and $t_j$
are fixed.
Only
$q_j(\boldsymbol{\phi}_{\mathrm{p}})
=\boldsymbol{\theta}_{\mathrm{c}}^{T}
\mathbf{M}_{\mathrm{e},j}\boldsymbol{\theta}_{\mathrm{c}}^{\ast}$
depends on $\boldsymbol{\phi}_{\mathrm{p}}$.
The UL branch power is therefore
$P_{\mathrm{s}}(\boldsymbol{\phi}_{\mathrm{p}})
=\sum_{j=\text{1}}^{J}|\varrho_{\mathrm{e},j}|^{\text{2}}r_j
[q_j(\boldsymbol{\phi}_{\mathrm{p}})
+\delta_{\mathrm{a}}^{\text{2}}t_j]
+\delta_{\mathrm{a}}^{\text{2}}
\|\boldsymbol{\Phi}_{\mathrm{a}}^{\mathrm{s}}\|_{F}^{\text{2}}$.
The corresponding HRIS and total powers,
$P_{\mathrm{ris}}(\boldsymbol{\phi}_{\mathrm{p}})
=P_{\mathrm{sta}}+P_{\mathrm{c}}
+P_{\mathrm{s}}(\boldsymbol{\phi}_{\mathrm{p}})$
and
$P_{\mathrm{tot}}(\boldsymbol{\phi}_{\mathrm{p}})
=\mathrm{Tr}(\mathbf{R}_{\mathrm{x}})+P_{\mathrm{f}}
+P_{\mathrm{ris}}(\boldsymbol{\phi}_{\mathrm{p}})$,
are convex quadratic functions of $\boldsymbol{\phi}_{\mathrm{p}}$.

Next, we address the rate constraints
\eqref{eq:interference_split} and \eqref{eq:signal_split1}.
With
$\Delta\boldsymbol{\phi}_{\mathrm{p}}
=\boldsymbol{\phi}_{\mathrm{p}}
-\boldsymbol{\phi}_{\mathrm{p}}^{(t)}$,
the lower bounds of user-signal and
Eve-interference terms are
\begin{subequations}
\label{eq:passive_rate_lower_bounds}
\begin{align}
\underline{D}_{\mathrm{u},k}^{\mathrm{p},(t)}
={}&
D_{\mathrm{u},k}(\boldsymbol{\phi}_{\mathrm{p}}^{(t)})
+\text{2}\mathrm{Re}\left\{
\boldsymbol{\theta}_{\mathrm{c}}^{(t),T}
\mathbf{Z}_{\mathrm{u},k}\mathbf{B}_{\mathrm{p}}^{\ast}
\Delta\boldsymbol{\phi}_{\mathrm{p}}^{\ast}
\right\},
\label{eq:passive_user_lb}\\
\underline{I}_{\mathrm{e},j,k}^{\mathrm{p},(t)}
={}&
I_{\mathrm{e},j,k}(\boldsymbol{\phi}_{\mathrm{p}}^{(t)})
+\text{2}\mathrm{Re}\left\{
\boldsymbol{\theta}_{\mathrm{c}}^{(t),T}
\mathbf{Z}_{\mathrm{e},j,-k}\mathbf{B}_{\mathrm{p}}^{\ast}
\Delta\boldsymbol{\phi}_{\mathrm{p}}^{\ast}
\right\}.
\label{eq:passive_eve_lb}
\end{align}
\end{subequations}
Then, the rate constraints are
conservatively approximated by
\begin{subequations}
\label{eq:passive_comm}
\begin{alignat}{2}
&\underline{D}_{\mathrm{u},k}^{\mathrm{p},(t)}
\geq g_{\mathrm{u},k}^{\mathrm{ub},(t)},
&&D_{\mathrm{e},j,k}(\boldsymbol{\phi}_{\mathrm{p}})
\leq g_{\mathrm{e},j,k}^{\mathrm{lb},(t)},
\forall j\in\mathcal{J},\forall k\in\mathcal{K},
\label{eq:passive_signal_constraints}\displaybreak[1]\\ 
&I_{\mathrm{u},k}(\boldsymbol{\phi}_{\mathrm{p}})
\leq\varpi_{\mathrm{u},k},
&&\underline{I}_{\mathrm{e},j,k}^{\mathrm{p},(t)}
\geq\varpi_{\mathrm{e},j,k},
\forall j\in\mathcal{J},\forall k\in\mathcal{K}.
\label{eq:passive_interference_constraints}
\end{alignat}
\end{subequations}

For sensing, recall
$\mathbf{H}_{\nu}^{\prime}
=\bar{\boldsymbol{G}}\boldsymbol{\Theta}_{\mathrm{s}}
\mathbf{A}_{\nu}\boldsymbol{\Theta}_{\mathrm{c}}\boldsymbol{G}$,
where $\mathbf{A}_{\nu}$ denotes the derivative matrices in \eqref{eq:CRB_unified representation} and is fixed in this update.
The passive phase vector enters both
$\boldsymbol{\Theta}_{\mathrm{c}}$ and
$\boldsymbol{\Theta}_{\mathrm{s}}$, making
$\mathbf{H}_{\nu}^{\prime}$ quadratic in
$\boldsymbol{\phi}_{\mathrm{p}}$.
To construct a local affine FIM approximation, define
$\boldsymbol{\xi}_{\mathrm{p}}
=[\mathrm{Re}\{\boldsymbol{\phi}_{\mathrm{p}}\}^{T},
\mathrm{Im}\{\boldsymbol{\phi}_{\mathrm{p}}\}^{T}]^{T}$.
At the reference point $\boldsymbol{\xi}_{\mathrm{p}}^{(t)}$,
the approximation is
\begin{equation}
\widehat{\mathbf{F}}_{\mathrm{p}}^{(t)}
=\mathbf{F}(\boldsymbol{\xi}_{\mathrm{p}}^{(t)})
+\sum_{l=\text{1}}^{\text{2}M}
\mathbf{F}_{l}^{(t)}
(\xi_{\mathrm{p},l}-\xi_{\mathrm{p},l}^{(t)}),
\label{eq:passive_FIM_lower}
\end{equation}
where
$\mathbf{F}_{l}^{(t)}
=\left.
\partial\mathbf{F}/\partial\xi_{\mathrm{p},l}
\right|_{\boldsymbol{\xi}_{\mathrm{p}}
=\boldsymbol{\xi}_{\mathrm{p}}^{(t)}}$. Then, we evaluate these derivative matrices. Only the derivatives of $\mathbf{U}$ are needed due to $\mathbf{Y}$ is independent of $\boldsymbol{\phi}_{\mathrm{p}}$.
Since the passive response is shared by both branches,
$\boldsymbol{\Theta}_{l}^{\prime}
\triangleq
\partial\boldsymbol{\Theta}_{\mathrm{c}}/\partial\xi_{\mathrm{p},l}
=\partial\boldsymbol{\Theta}_{\mathrm{s}}/\partial\xi_{\mathrm{p},l}
=\mathrm{diag}(\mathbf{B}_{\mathrm{p}}
\bar{\boldsymbol{\epsilon}}_l)$,
where $\bar{\boldsymbol{\epsilon}}_l=\boldsymbol{\epsilon}_l$
for $l\leq M$ and
$\bar{\boldsymbol{\epsilon}}_l
=\mathrm{j}\boldsymbol{\epsilon}_{l-M}$ otherwise,
with $\boldsymbol{\epsilon}_m$ denoting the $m$-th column of
$\mathbf{I}_M$. Then, we have $\frac{\partial\mathbf{H}_{\nu}^{\prime}}
{\partial\xi_{\mathrm{p},l}}
=\bar{\boldsymbol{G}}\left(
\boldsymbol{\Theta}_{l}^{\prime}\mathbf{A}_{\nu}
\boldsymbol{\Theta}_{\mathrm{c}}
+\boldsymbol{\Theta}_{\mathrm{s}}\mathbf{A}_{\nu}
\boldsymbol{\Theta}_{l}^{\prime}
\right)\boldsymbol{G}$.
Using
$\mathbf{U}_{l}^{\prime}
=\partial\mathbf{U}/\partial\xi_{\mathrm{p},l}$,
the FIM derivative at the reference point is
\begin{equation}
\mathbf{F}_{l}^{(t)}
=\text{2}L\mathrm{Re}\left\{
\mathbf{U}_{l}^{\prime,(t),H}
\mathbf{Y}^{-\text{1}}\mathbf{U}^{(t)}
+\mathbf{U}^{(t),H}\mathbf{Y}^{-\text{1}}
\mathbf{U}_{l}^{\prime,(t)}
\right\}.
\end{equation}

Substituting \eqref{eq:passive_FIM_lower} into
\eqref{eq:crb_schur_lmi} gives a linear matrix inequality.
The convex PCCP subproblem for
$\boldsymbol{\phi}_{\mathrm{p}}$ is given by
\begin{subequations}
\label{eq:P13}
\begin{eqnarray}
\mathrm{P\text{13}}:
&\underset{\boldsymbol{\phi}_{\mathrm{p}},\boldsymbol{\Delta}}{\max}
&
\psi-\tau_{\mathrm{p}}
\overline{\mathcal{R}}_{\phi}^{(t)}
(\boldsymbol{\phi}_{\mathrm{p}})
\label{eq:P13_obj}\\
&\mathrm{s.t.}&
\eqref{eq:P2_qos}-\eqref{eq:P2_positive},
\eqref{eq:crb_trace_aux},\eqref{eq:passive_comm},
\label{eq:P13_1}\\
&&
P_{\mathrm{ris}}(\boldsymbol{\phi}_{\mathrm{p}})
\leq P_{\mathrm{ris},\max},
\label{eq:P13_hris_power}\\
&&
P_{\mathrm{tot}}(\boldsymbol{\phi}_{\mathrm{p}})
\leq\eta,
\label{eq:P13_total_power}\\
&&
\begin{bmatrix}
\widehat{\mathbf{F}}_{\mathrm{p}}^{(t)}
&\mathbf{J}_j^{T}\\
\mathbf{J}_j&\mathbf{C}_j
\end{bmatrix}
\succeq\boldsymbol{\text{0}},
\ \forall j\in\mathcal{J},
\label{eq:P13_crb}\\
&&
|[\boldsymbol{\phi}_{\mathrm{p}}]_m|^{\text{2}}
\leq\text{1},
\ \forall m\in\mathcal{M}.
\label{eq:P13_phase}
\end{eqnarray}
\end{subequations}
Here, $\tau_{\mathrm{p}}>\text{0}$ is the passive phase penalty weight.

%%%%%%%%%%%%%%%%%%%%%%%%%%%%%%%%%%%%%%%%%%%%\vspace{-4mm}
\subsection{Active-Element Selection Optimization}

With the continuous state $\mathcal{X}$ fixed, the binary active-element configuration subproblem is given by
\begin{subequations}
\label{eq:p14_binary}
\begin{eqnarray}
\mathrm{P\text{14}}:
&\underset{\boldsymbol{e}\in\{\text{0},\text{1}\}^{M},
\boldsymbol{\Delta}}{\max}
&\psi
\label{eq:p14_obj}\\
&\mathrm{s.t.}&
\eqref{eq:P0_hris_power},\eqref{eq:P1_power},
\eqref{eq:P2_qos}-\eqref{eq:P2_positive},
\label{eq:p14_1}\\
&&
\eqref{eq:interference_split},\eqref{eq:signal_split1},
\eqref{eq:crb_schur}.
\label{eq:p14_2}
\end{eqnarray}
\end{subequations}

To generate candidate configurations without enumerating all
$\text{2}^{M}$ binary patterns, we design a full-aperture
reweighted group-sparse selection method~\cite{r28}, which jointly updates configuration and HRIS responses with BS beamforming fixed. Since the $m$-th entries of the DL and UL active-response
vectors share the same selection indicator $e_m$,
we group them as
$\boldsymbol{g}_m
=[\bar{\phi}_{\mathrm{a},m}^{\mathrm{c}},
\bar{\phi}_{\mathrm{a},m}^{\mathrm{s}}]^T$.
We relax the binary indicators to continuous activation
variables $\varsigma_m\in[\text{0},\text{1}]$ and impose
\begin{equation}
\text{0}\leq \varsigma_m\leq\text{1},\ 
\left\Vert\boldsymbol{g}_m\right\Vert_{\text{2}}\leq\alpha_{\max}\varsigma_m,\ 
\left|\bar{\phi}_{\mathrm{p},m}\right|\leq\text{1}-\varsigma_m,\forall m \in \mathcal{M} .
\label{eq:p14_gate}
\end{equation}

For the power constraints, both active response vectors vary,
making the products $r_jq_j$ and $r_jt_j$ in
\eqref{eq:branch_echo_output} non-convex.
Therefore, we introduce
nonnegative auxiliary upper bounds
\begin{subequations}
\label{eq:r_t_j}
\begin{align}
\bar{r}_j
&\geq r_j
=(\bar{\boldsymbol{\phi}}_{\mathrm{a}}^{\mathrm{s}})^{H}
\mathrm{diag}(|\boldsymbol{h}_{\mathrm{e},j}|^{\text{2}})
\bar{\boldsymbol{\phi}}_{\mathrm{a}}^{\mathrm{s}},\\
\bar{q}_j
&\geq q_j
=\boldsymbol{\theta}_{\mathrm{c}}^{T}
\mathbf{M}_{\mathrm{e},j}\boldsymbol{\theta}_{\mathrm{c}}^{\ast},\\
\bar{t}_j
&\geq t_j
=(\bar{\boldsymbol{\phi}}_{\mathrm{a}}^{\mathrm{c}})^{H}
\mathrm{diag}(|\boldsymbol{h}_{\mathrm{e},j}|^{\text{2}})
\bar{\boldsymbol{\phi}}_{\mathrm{a}}^{\mathrm{c}},
\ \forall j\in\mathcal{J}.
\end{align}
\end{subequations}
With
$\varphi_m
=[\boldsymbol{G}\mathbf{R}_{\mathrm{x}}^{(t)}
\boldsymbol{G}^{H}]_{m,m}$, the convex upper bound on the relaxed HRIS power is written as
\begin{align}
\widehat{P}_{\mathrm{ris}}^{(t)}
={}&MP_{\mathrm{p}}
+(P_{\mathrm{a}}-P_{\mathrm{p}})
\mathbf{\text{1}}_{M}^{T}\boldsymbol{\varsigma}
+\sum_{m=\text{1}}^{M}
(\varphi_m+\delta_{\mathrm{a}}^{\text{2}})
|\bar{\phi}_{\mathrm{a},m}^{\mathrm{c}}|^{\text{2}}
\nonumber\\
&+\delta_{\mathrm{a}}^{\text{2}}
\|\bar{\boldsymbol{\phi}}_{\mathrm{a}}^{\mathrm{s}}\|_{\text{2}}^{\text{2}}
+\sum_{j=\text{1}}^{J}|\varrho_{\mathrm{e},j}|^{\text{2}}
\left[
\frac{\xi_{rq,j}^{(t)}}{\text{2}}\bar{r}_j^{\text{2}}
+\frac{\bar{q}_j^{\text{2}}}{\text{2}\xi_{rq,j}^{(t)}}
\right.
\nonumber\\
&\left.
+\delta_{\mathrm{a}}^{\text{2}}\left(
\frac{\xi_{rt,j}^{(t)}}{\text{2}}\bar{r}_j^{\text{2}}
+\frac{\bar{t}_j^{\text{2}}}{\text{2}\xi_{rt,j}^{(t)}}
\right)
\right],
\label{eq:p14_power_bound}
\end{align}
where
$\xi_{rq,j}^{(t)}=q_j^{(t)}/r_j^{(t)}$
and
$\xi_{rt,j}^{(t)}=t_j^{(t)}/r_j^{(t)}$
for positive reference values
$r_j^{(t)}$, $q_j^{(t)}$, and $t_j^{(t)}$.

Next, we address the rate constraints
\eqref{eq:interference_split} and \eqref{eq:signal_split1}.
The signal power is convex quadratic in
$\boldsymbol{\theta}_{\mathrm{c}}$, and the interference-plus-noise
power is convex quadratic in
$\left(\boldsymbol{\theta}_{\mathrm{c}},
\bar{\boldsymbol{\phi}}_{\mathrm{a}}^{\mathrm{c}}\right)$.
Let
$\Delta\boldsymbol{\theta}_{\mathrm{c}}
=\boldsymbol{\theta}_{\mathrm{c}}
-\boldsymbol{\theta}_{\mathrm{c}}^{(t)}$
and
$\Delta\bar{\boldsymbol{\phi}}_{\mathrm{a}}^{\mathrm{c}}
=\bar{\boldsymbol{\phi}}_{\mathrm{a}}^{\mathrm{c}}
-\bar{\boldsymbol{\phi}}_{\mathrm{a}}^{\mathrm{c},(t)}$.
The lower bounds of user-signal and Eve-interference terms are
\begin{subequations}
\label{eq:p14_rate_lower_bounds}
\begin{align}
\underline{D}_{\mathrm{u},k}^{\mathrm{e},(t)}
={}&D_{\mathrm{u},k}^{(t)}
+\text{2}\mathrm{Re}\left\{
\boldsymbol{\theta}_{\mathrm{c}}^{(t),T}
\mathbf{Z}_{\mathrm{u},k}
(\Delta\boldsymbol{\theta}_{\mathrm{c}})^{\ast}
\right\},
\label{eq:p14_user_signal_lb}\\
\underline{I}_{\mathrm{e},j,k}^{\mathrm{e},(t)}
={}&I_{\mathrm{e},j,k}^{(t)}
+\text{2}\mathrm{Re}\left\{
\boldsymbol{\theta}_{\mathrm{c}}^{(t),T}
\mathbf{Z}_{\mathrm{e},j,-k}
(\Delta\boldsymbol{\theta}_{\mathrm{c}})^{\ast}
\right\}
\nonumber\\
&+\text{2}\delta_{\mathrm{a}}^{\text{2}}
\mathrm{Re}\left\{
(\bar{\boldsymbol{\phi}}_{\mathrm{a}}^{\mathrm{c},(t)})^{H}
\mathrm{diag}(|\boldsymbol{h}_{\mathrm{e},j}|^{\text{2}})
\Delta\bar{\boldsymbol{\phi}}_{\mathrm{a}}^{\mathrm{c}}
\right\},
\label{eq:p14_eve_interference_lb}
\end{align}
\end{subequations}
where $D_{\mathrm{u},k}^{(t)}$ and
$I_{\mathrm{e},j,k}^{(t)}$ denote the powers evaluated at
the reference responses.
Therefore, the rate constraints are conservatively approximated by
\begin{subequations}
\label{eq:p14_comm}
\begin{alignat}{2}
&\underline{D}_{\mathrm{u},k}^{\mathrm{e},(t)}
\geq g_{\mathrm{u},k}^{\mathrm{ub},(t)},\ 
&&D_{\mathrm{e},j,k}
\leq g_{\mathrm{e},j,k}^{\mathrm{lb},(t)},
\forall j\in\mathcal{J},\forall k\in\mathcal{K},
\label{eq:p14_signal_constraints}\\
&I_{\mathrm{u},k}\leq\varpi_{\mathrm{u},k},
&&\underline{I}_{\mathrm{e},j,k}^{\mathrm{e},(t)}
\geq\varpi_{\mathrm{e},j,k},
\forall j\in\mathcal{J},\forall k\in\mathcal{K}.
\label{eq:p14_interference_constraints}
\end{alignat}
\end{subequations}

For sensing, the DL and UL responses vary jointly, so
$\boldsymbol{\Theta}_{\mathrm{s}}\mathbf{A}_{\nu}
\boldsymbol{\Theta}_{\mathrm{c}}$
in $\mathbf{H}_{\nu}^{\prime}$ is bilinear.
We use its local first-order approximation $\widehat{\mathbf{H}}_{\nu}^{\prime,(t)}
=\bar{\boldsymbol{G}}\left(
\boldsymbol{\Theta}_{\mathrm{s}}^{(t)}
\mathbf{A}_{\nu}\boldsymbol{\Theta}_{\mathrm{c}}
+\boldsymbol{\Theta}_{\mathrm{s}}\mathbf{A}_{\nu}
\boldsymbol{\Theta}_{\mathrm{c}}^{(t)}
-\boldsymbol{\Theta}_{\mathrm{s}}^{(t)}
\mathbf{A}_{\nu}\boldsymbol{\Theta}_{\mathrm{c}}^{(t)}
\right)\boldsymbol{G}$. 
Define
$\widehat{\mathbf{U}}^{(t)}
=\left[
\mathrm{vec}\left(
\widehat{\mathbf{H}}_{\nu}^{\prime,(t)}
(\mathbf{R}_{\mathrm{x}}^{(t)})^{\text{1}/\text{2}}
\right)
\right]_{\nu=\text{1}}^{\text{5}J}$,
which is affine in the relaxed HRIS responses.
To construct an affine upper bound on $\mathbf{R}_{\omega}$, we introduce
$\boldsymbol{z}=[z_{\text{1}},\dots,z_M]^T$
and replace the quadratic terms
$|\bar{\phi}_{\mathrm{a},m}^{\mathrm{s}}|^{\text{2}}$
with auxiliary upper bounds
$z_m\geq|\bar{\phi}_{\mathrm{a},m}^{\mathrm{s}}|^{\text{2}}$.
This gives
$\overline{\mathbf{Y}}(\boldsymbol{z})
=\mathbf{I}_{N}\otimes
[\delta_{\mathrm{a}}^{\text{2}}
\bar{\boldsymbol{G}}\mathrm{diag}(\boldsymbol{z})
\bar{\boldsymbol{G}}^{H}+\mathbf{R}_{\mathrm{BR}}]$.
Applying \eqref{eq:branch_fim_lower} with
$\widehat{\mathbf{U}}^{(t)}$ and
$\overline{\mathbf{Y}}(\boldsymbol{z})$ gives
$\underline{\mathbf{F}}_{\mathrm{e}}^{(t)}
=\underline{\mathbf{F}}^{(t)}
(\widehat{\mathbf{U}}^{(t)},
\overline{\mathbf{Y}}(\boldsymbol{z}))$,
an affine FIM lower bound for the locally linearized echo model.
The reference covariance is evaluated at
$z_m^{(t)}
=|\bar{\phi}_{\mathrm{a},m}^{\mathrm{s},(t)}|^{\text{2}}$.

At reweighting iteration $l$, let $\kappa_m^{(l)}>\text{0}$
denote the element weights and $\tau_{\mathrm{e}}>\text{0}$
the sparsity penalty weight.
The convex surrogate used to obtain continuous solutions is given by
\begin{subequations}
\label{eq:p15_reweighted}
\begin{eqnarray}
\mathrm{P\text{15}}:
&\underset{\substack{
\boldsymbol{\varsigma},
\bar{\boldsymbol{\phi}}_{\mathrm{a}}^{\mathrm{c}},
\bar{\boldsymbol{\phi}}_{\mathrm{a}}^{\mathrm{s}},
\bar{\boldsymbol{\phi}}_{\mathrm{p}},\\
\boldsymbol{z},
\{\bar{r}_j,\bar{q}_j,\bar{t}_j\},
\boldsymbol{\Delta}
}}{\max}
&
\psi-\tau_{\mathrm{e}}
\sum_{m=\text{1}}^{M}\kappa_m^{(l)}\varsigma_m
\label{eq:p15_obj}\\
&\mathrm{s.t.}&
\eqref{eq:P2_qos}-\eqref{eq:P2_positive},
\eqref{eq:crb_trace_aux},\\
&&
\eqref{eq:p14_gate},\eqref{eq:r_t_j},\eqref{eq:p14_comm},
\label{eq:p15_rate_constraints}\\
&&
\widehat{P}_{\mathrm{ris}}^{(t)}
\leq P_{\mathrm{ris},\max},
\label{eq:p14_hris_power}\\
&&
\mathrm{Tr}(\mathbf{R}_{\mathrm{x}}^{(t)})
+P_{\mathrm{f}}+\widehat{P}_{\mathrm{ris}}^{(t)}
\leq\eta,
\label{eq:p14_total_power}\\
&&
\begin{bmatrix}
\underline{\mathbf{F}}_{\mathrm{e}}^{(t)}
&\mathbf{J}_j^{T}\\
\mathbf{J}_j&\mathbf{C}_j
\end{bmatrix}
\succeq\boldsymbol{\text{0}},
\forall j\in\mathcal{J},
\label{eq:p14_crb_lmi}\\
&&
\begin{bmatrix}
z_m&\bar{\phi}_{\mathrm{a},m}^{\mathrm{s}}\\
\bar{\phi}_{\mathrm{a},m}^{\mathrm{s},\ast}&\text{1}
\end{bmatrix}
\succeq\boldsymbol{\text{0}},
\forall m\in\mathcal{M}.
\label{eq:p14_z}
\end{eqnarray}
\end{subequations}
The resulting activation values $\varsigma_m^{(l)}$
are used to update and normalize the weights as $\kappa_m^{(l+\text{1})}
=\widetilde{\kappa}_m^{(l+\text{1})}/\left(
M^{-\text{1}}\sum_{n=\text{1}}^{M}
\widetilde{\kappa}_n^{(l+\text{1})}\right)$,
where
$\widetilde{\kappa}_m^{(l+\text{1})}
=\min\{\text{1}/(\varsigma_m^{(l)}+\varepsilon_s),
\kappa_{\max}\}$,
$\varepsilon_s>\text{0}$ is a regularization parameter,
and $\kappa_{\max}>\text{0}$ caps the unnormalized weights.
This reweighting assigns larger penalties to weakly activated
elements, promoting sparsity across the response groups
$\{\boldsymbol{g}_m\}_{m=1}^{M}$ for subsequent binary recovery.

After reweighting, we calculate the selection score
$\omega_m
=\beta\varsigma_m
+(\text{1}-\beta)
\min\{\|\boldsymbol{g}_m\|_{\text{2}}/\alpha_{\max},\text{1}\}$
from the continuous solution of $\mathrm{P\text{15}}$ and rank the HRIS elements
in descending order.
Here, $\beta\in[\text{0},\text{1}]$ balances the activation
level and normalized active-response magnitude.
To determine the active-element number, we  consider candidate
counts satisfying the necessary static-power condition,
i.e.,
$s\in\mathcal{S}^{(t)}
\subseteq
\{s\in\{\text{0},\ldots,M\}
\mid sP_{\mathrm{a}}+(M-s)P_{\mathrm{p}}
\leq P_{\mathrm{ris},\max}\}$,where $\mathcal{S}^{(t)}$ is constructed through global sampling
or local sampling around the current count.
For each count $s$, we recover a binary configuration
by activating the $s$ highest-scoring elements.
For each recovered configuration, we refine the continuous
HRIS variables
$\mathcal{X}_{\mathrm{e}}
\triangleq\mathcal{X}
\setminus\{\mathbf{W}_{\mathrm{c}},\mathbf{W}_{\mathrm{s}}\}$
with the BS beamformers fixed.
We then verify the original constraints of $\mathrm{P\text{0}}$,
including the CRB requirement, together with
$R_{\mathrm{u},k}
\geq\max_{j\in\mathcal{J}}R_{\mathrm{e},j,k}$,
$\forall k\in\mathcal{K}$.
Among feasible candidates, we select the configuration with
the highest genuine SEE, denoted by $\mathrm{SEE}_{\mathrm{ou}}$.
The configuration update is accepted only if
$\mathrm{SEE}_{\mathrm{ou}}-\mathrm{SEE}_{\mathrm{in}}
\geq\tau_{\mathrm{acc}}
\max\{\text{1},|\mathrm{SEE}_{\mathrm{in}}|\}$,
where $\tau_{\mathrm{acc}}>\text{0}$ and
$\mathrm{SEE}_{\mathrm{in}}$ is the genuine SEE before
the selection update.
Upon acceptance, both the configuration and its associated
continuous HRIS variables are updated.

\vspace{-3mm}
\subsection{Complexity and Convergence Analysis}

Algorithms~\ref{alg:reweighted_selection} and
\ref{alg:overall_ao} summarize the selection and proposed AO procedures.
Using interior-point methods~\cite{r27}, the per-solve costs are
$S_{\text{5}}
=\mathcal{O}((NK+N^{\text{2}})^{\text{3.5}}
+J(\text{5}J+\text{3})^{\text{3}})$
for BS beamforming and
$S_{\mathrm{h}}
=\mathcal{O}(M^{\text{3}}+J(\text{5}J+\text{3})^{\text{3}})$
for HRIS updates.
With $I_{\mathrm{rew}}$ reweighting iterations and per-candidate
HRIS refinement cost $S_{\mathcal{X}_{\mathrm{e}}}$,
selection costs
$S_{\mathrm{sel}}^{(t)}
=\mathcal{O}(I_{\mathrm{rew}}S_{\mathrm{h}}
+M\log M+|\mathcal{S}^{(t)}|S_{\mathcal{X}_{\mathrm{e}}})$.
The overall complexity is
$\mathcal{O}\big(I_{\mathrm{AO}}\big[
S_{\text{5}}+(\text{2}+\text{3}I_{\mathrm{PCCP}})S_{\mathrm{h}}
+\max_t S_{\mathrm{sel}}^{(t)}\big]\big)$,
where $I_{\mathrm{AO}}$ and $I_{\mathrm{PCCP}}$
count outer AO and inner PCCP iterations. With refreshed auxiliary variables, tight conservative SCA
bounds ensure non-decreasing genuine SEE for the beamforming
and DL gain updates; the UL gain update minimizes power
at fixed secrecy rate.
Phase and configuration updates are accepted only if they
satisfy the original constraints without decreasing genuine SEE.
The resulting SEE sequence is non-decreasing and bounded above
under finite power and gain budgets and positive circuit power,
ensuring convergence of the objective value.

\begin{algorithm}[t]
\caption{Reweighted Group-Sparse Selection for Active-Element Configuration}
\label{alg:reweighted_selection}
\begin{algorithmic}[1]
\STATE Initialize $\kappa_m^{(\text{0})}=\text{1}$ and $l=\text{0}$; set $\epsilon_{\mathrm{rew}}$, selection parameters, and $\mathcal{S}^{(t)}$; evaluate $\mathrm{SEE}_{\mathrm{in}}$ at the current $(\boldsymbol{e},\mathcal{X})$;
\REPEAT
    \STATE Solve $\mathrm{P\text{15}}$ using $\kappa_m^{(l)}$, obtaining $\boldsymbol{\varsigma}^{(l)}$ and HRIS responses;
    \STATE Update $\kappa_m^{(l+\text{1})}$ via $\kappa_m^{(l+\text{1})}=\widetilde{\kappa}_m^{(l+\text{1})}/\left(M^{-\text{1}}\sum_{n=\text{1}}^{M}\widetilde{\kappa}_n^{(l+\text{1})}\right)$;
    \STATE $l\leftarrow l+\text{1}$;
\UNTIL{the change in $\boldsymbol{\varsigma}$ is at most $\epsilon_{\mathrm{rew}}$}
\STATE Calculate selection scores $\omega_m$, $\forall m\in\mathcal{M}$;
\FOR{each $s\in\mathcal{S}^{(t)}$}
    \STATE Activate the $s$ highest-scoring elements to form a binary configuration;
    \STATE Refine $\mathcal{X}_{\mathrm{e}}$ with BS beamforming fixed and evaluate genuine SEE for feasible candidates;
\ENDFOR
\STATE Identify $(\boldsymbol{e}_{\mathrm{cand}},\mathcal{X}_{\mathrm{e,cand}})$ with the highest feasible genuine SEE, denoted by $\mathrm{SEE}_{\mathrm{ou}}$;
\IF{$\mathrm{SEE}_{\mathrm{ou}}-\mathrm{SEE}_{\mathrm{in}}\geq\tau_{\mathrm{acc}}\max\{\text{1},|\mathrm{SEE}_{\mathrm{in}}|\}$}
    \STATE Update $\boldsymbol{e}\leftarrow\boldsymbol{e}_{\mathrm{cand}}$ and $\mathcal{X}_{\mathrm{e}}\leftarrow\mathcal{X}_{\mathrm{e,cand}}$;
\ENDIF
\end{algorithmic}
\end{algorithm}

\vspace{-2mm}
\begin{algorithm}[t]
\caption{Proposed AO Framework for Problem $\mathrm{P\text{0}}$}
\label{alg:overall_ao}
\begin{algorithmic}[1]
\STATE Initialize $\mathcal{X}$ and $\boldsymbol{e}$; set PCCP parameters $\{c>\text{1},\tau_{\max},\varepsilon_{\phi,\text{1}},\varepsilon_{\phi,\text{2}}\}$ and $\epsilon_{\mathrm{AO}}$; evaluate $\mathrm{SEE}^{(\text{0})}$; $t\leftarrow\text{0}$;
\REPEAT
    \STATE Update $\{\mathbf{W}_{\mathrm{c}},\mathbf{W}_{\mathrm{s}}\}$ by solving $\mathrm{P\text{5}}$;
    \STATE Update $\boldsymbol{\alpha}$ and $\boldsymbol{\rho}$ by solving $\mathrm{P\text{7}}$ and $\mathrm{P\text{8}}$;
    \FOR{each $\boldsymbol{\phi}\in\{\boldsymbol{\phi}_{\mathrm{a}}^{\mathrm{c}},\boldsymbol{\phi}_{\mathrm{a}}^{\mathrm{s}},\boldsymbol{\phi}_{\mathrm{p}}\}$}
        \STATE Initialize $\boldsymbol{\phi}^{(\text{0})}\leftarrow\boldsymbol{\phi}$, $\tau\leftarrow\tau^{(\text{0})}$, and $i=\text{0}$;
        \REPEAT
            \STATE Obtain $\boldsymbol{\phi}^{(i+\text{1})}$ by solving $\mathrm{P\text{10}}$, $\mathrm{P\text{11}}$, or $\mathrm{P\text{13}}$;
            \STATE $\tau\leftarrow\min\{c\tau,\tau_{\max}\}$ and $i\leftarrow i+\text{1}$;
        \UNTIL{$\|\boldsymbol{\phi}^{(i)} \!- \!\boldsymbol{\phi}^{(i-\text{1})}\|_{\text{2}} \!\leq \!\varepsilon_{\phi,\text{1}}$ \textbf{and} $\mathcal{R}_{\phi}(\boldsymbol{\phi}^{(i)}) \!\leq\! \varepsilon_{\phi,\text{2}}$}
        \STATE Recover the unit-modulus candidate and update $\boldsymbol{\phi}$ if feasible and genuine SEE does not decrease;
    \ENDFOR
    \STATE Update $\boldsymbol{e}$ and $\mathcal{X}_{\mathrm{e}}$ via \textbf{Algorithm~\ref{alg:reweighted_selection}};
    \STATE Evaluate $\mathrm{SEE}^{(t+\text{1})}$ at the current $(\boldsymbol{e},\mathcal{X})$;
    \STATE $t\leftarrow t+\text{1}$;
\UNTIL{$|\mathrm{SEE}^{(t)}-\mathrm{SEE}^{(t-\text{1})}|/\mathrm{SEE}^{(t-\text{1})}\leq\epsilon_{\mathrm{AO}}$}
\end{algorithmic}
\end{algorithm}

%\vspace{-2mm}
\section{Simulation Results}
In this section, numerical simulations are provided to evaluate the proposed HRIS-assisted secure near-field ISAC scheme. Unless otherwise specified, the default simulation parameters are configured as follows. We consider a representative high-frequency near-field scenario with a carrier frequency of $f_c = \text{220}$~GHz~\cite{r_frenquency} with an equivalent noise bandwidth of $\text{100}$~MHz. The HRIS is centered at $(\text{0}, \text{0}, \text{0})$~m and comprises $M = \text{40} \times \text{40}$ elements. The BS, located at $(\text{0.6}, -\text{0.3}, \text{0.2})$~m, employs $N = \text{16}$ shared antennas to serve $K = \text{2}$ users and sense $J = \text{2}$ potential Eves. The distances from the HRIS to the users and Eves are in $[\text{0.4}, \text{1.8}]$~m, and the azimuth-angle separation between each user and its corresponding Eve is $12^\circ$.  The Rician factors are set to $\kappa_{\mathrm{b}} = \text{15}$~dB and $\kappa_{p,q} = \text{10}$~dB, with distance-independent link power gains $\beta_{\mathrm{b}}=\beta_{p,q}=\text{10}$. The molecular absorption follows the exponential attenuation model in~\cite{r_absorption}.
The receiver and amplification noise powers are set to $\delta_{p,q}^{\text{2}}=-\text{90}$~dBm, $\forall p,q$ and $\delta_{\mathrm{a}}^{\text{2}}=-\text{92}$~dBm, respectively. At the BS, the residual self-interference is modeled using a $\text{120}$~dB suppression factor~\cite{r_fullduplex} and a fixed $\text{30}$~dBm reference leakage power. The aggregate BS disturbance covariance is set to $\mathbf{R}_{\mathrm{BR}}=\left(\text{4.0124}\times\text{10}^{-\text{12}}\right)\mathbf{I}_{N}$~W. 
The power and hardware budgets are $P_{\max} = \text{21}$~dBm, $P_{\mathrm{ris},\max} = \text{30}$~dBm, $\alpha_{\max} = \text{3}$, $P_{\mathrm{f}} = \text{0.5}$~W, $P_{\mathrm{a}} = \text{2}$~mW, $P_{\mathrm{p}} = \text{0.1}$~mW, and $R_{\mathrm{u},\min}=\text{1}$~bit/s/Hz. Each CPI contains $L = \text{256}$ snapshots. The targets have a radar cross section of $\sigma_{\mathrm{e},j}=\text{1}\,\mathrm{m}^{\text{2}}$ and a zero nominal reflection phase, with $\left|\varrho_{\mathrm{e},j}\right|^{\text{2}}=\text{4}\pi\sigma_{\mathrm{e},j}/\lambda_c^{\text{2}}$. The CRB-based localization RMSE threshold is $\epsilon_{\max} = \text{5}$~cm.

\begin{figure}[t]
  \centering
\includegraphics[width=3.7in,trim=0 0 0 30, clip]{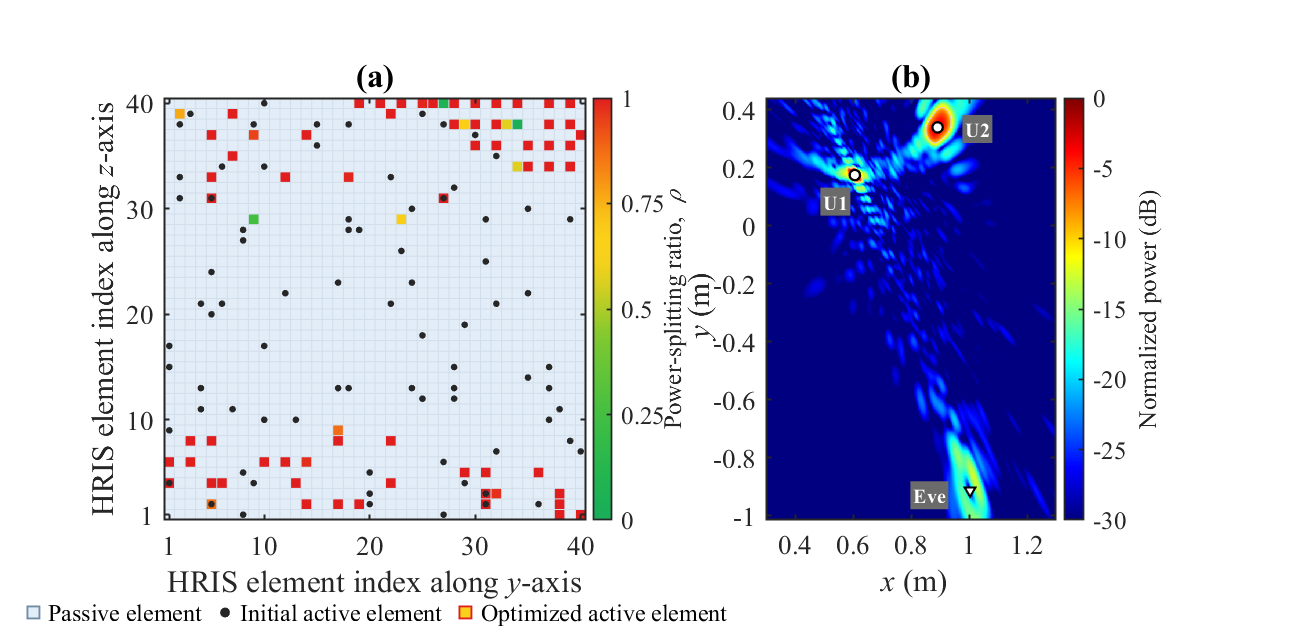}
\caption{Optimized HRIS configuration and beamforming behavior: (a) active-element distribution and power-splitting ratio $\rho$; (b) near-field spatial energy focusing on the terminal plane.}  \label{fig:fig_S1_Topology_and_Focusing}
\vspace{-4mm}
\end{figure}

Fig.~\ref{fig:fig_S1_Topology_and_Focusing} illustrates the optimized HRIS configuration and near-field energy distribution under $P_{\max} = \text{9}$~dBm, $\epsilon_{\max} = \text{1}$~cm, $K = \text{2}$, and $J = \text{1}$. In Fig.~\ref{fig:fig_S1_Topology_and_Focusing}(a), the proposed algorithm reduces the number of active elements from $\text{80}$ to $\text{73}$ and concentrates the active subset near the upper and lower edges. Moreover, most active elements operate with power-splitting ratios $\rho > \text{0.5}$, allocating a larger share of the amplification gain to the DL branch. In Fig.~\ref{fig:fig_S1_Topology_and_Focusing}(b), near-field beam focusing forms sharp energy peaks at legitimate users U1 and U2, while the total waveform power at Eve is $\text{30.6}$~dB below the spatial peak. These results illustrate the precise energy confinement and leakage suppression through joint element selection and beamforming.

\begin{figure}[!t]
\centering
\includegraphics[width=3.7in,trim=0 10 0 25, clip]{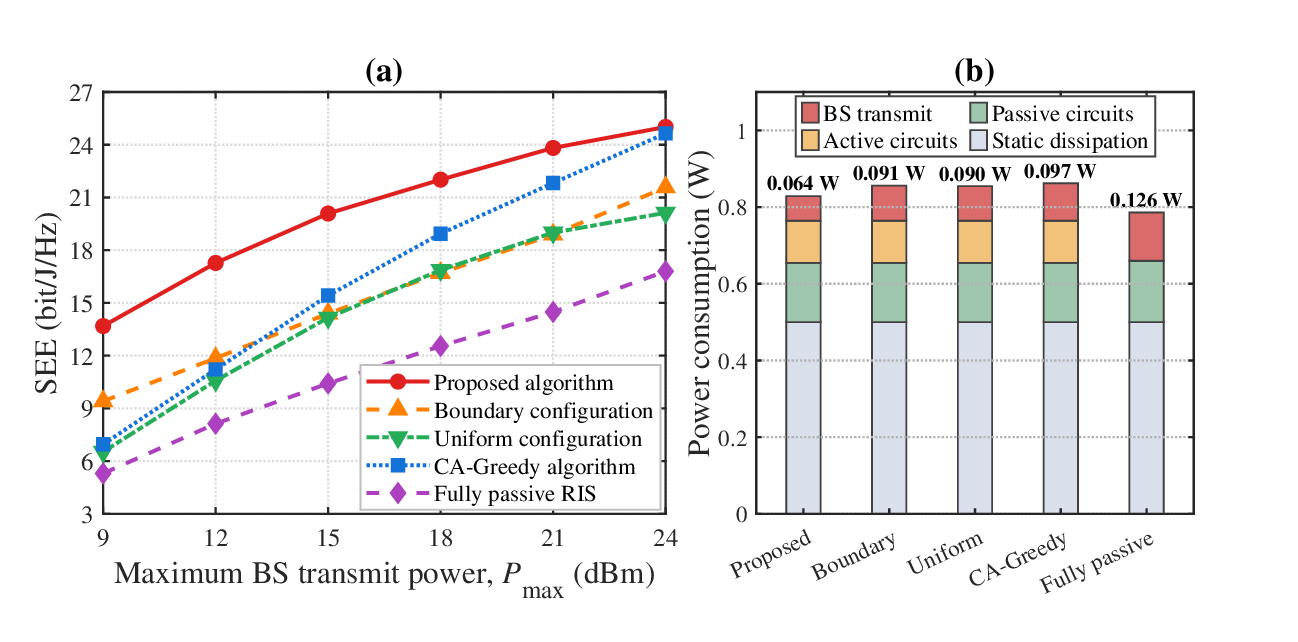}
\caption{SEE performance and power consumption breakdown: (a) SEE versus maximum BS transmit power $P_{\max}$; (b) power consumption breakdown of different schemes under $P_{\max} = \text{21}$~dBm.}
\label{fig_S2_transmit_power}
\vspace{-6mm}
\end{figure}

Fig.~\ref{fig_S2_transmit_power} investigates the system SEE and power consumption breakdown under various maximum BS transmit power budgets $P_{\max}$. We compare the proposed algorithm against four benchmark schemes: the boundary configuration with active elements placed along the outer aperture perimeter, the uniform configuration with active elements evenly spaced across the surface, the channel-aware greedy (CA-Greedy) scheme inspired by \cite{r19} that sequentially selects active elements according to the resulting SEE under fixed transmission parameters and subsequently optimizes the continuous variables for the selected configuration, and the fully passive RIS baseline. For fair comparison, both boundary and uniform benchmarks deploy the exact number of active elements optimized by the proposed algorithm under each parameter setup, ensuring identical active hardware budgets. In Fig.~\ref{fig_S2_transmit_power}(a), the SEE of all schemes increases monotonically with $P_{\max}$, because a higher power budget relaxes the transmit power constraint and provides greater beamforming flexibility. The proposed algorithm consistently achieves the highest SEE across the evaluated range. To evaluate the underlying power trade-off, Fig.~\ref{fig_S2_transmit_power}(b) illustrates the power consumption breakdown at $P_{\max} = \text{21}$~dBm, where the values above the bars indicate the BS transmit power $P_{\mathrm{BS}} = \operatorname{Tr}(\mathbf{R}_{\mathbf{x}})$. The proposed algorithm consumes approximately $\text{0.064}$~W of BS transmit power while satisfying both user rate and CRB sensing constraints, resulting in a total power consumption of $\text{0.829}$~W. Compared with CA-Greedy, the proposed algorithm reduces BS power by $\text{34.2}\%$ and total power consumption by $\text{3.9}\%$, while improving SEE by $\text{9.2}\%$. Although the fully passive RIS consumes the lowest total power ($\text{0.786}$~W), it expends nearly the entire BS transmit power budget ($\text{0.126}$~W) and still achieves the lowest SEE due to uncompensated cascaded path loss. These results suggest that the performance gain enabled by adaptive active elements effectively compensates for their additional circuit power cost.

\begin{figure}[!t]
\centering
\includegraphics[width=3.7in,trim=0 10 0 25, clip]{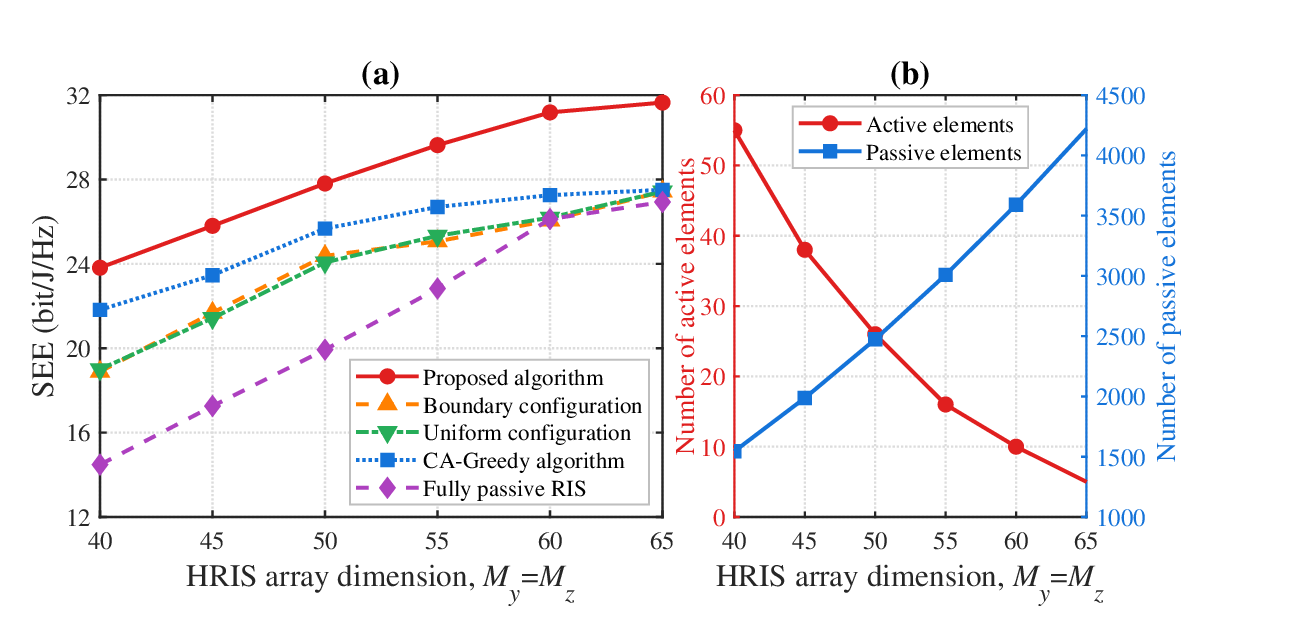}
\caption{SEE performance and element allocation versus the HRIS array dimension $M_y = M_z$: (a) SEE; (b) optimized numbers of active and passive elements.}
\label{fig_S3_HRIS_side}
\vspace{-5mm}
\end{figure}

Fig.~\ref{fig_S3_HRIS_side} investigates the system SEE and element allocation under various HRIS array dimensions $M_y = M_z \in [\text{40}, \text{65}]$. In Fig.~\ref{fig_S3_HRIS_side}(a), the SEE of all schemes increases monotonically with the array dimension, because a larger physical aperture sharpens near-field beam focusing. The proposed algorithm achieves the highest SEE across all aperture sizes. Meanwhile, the fully passive RIS improves faster than the HRIS-assisted schemes. Its performance gap with the proposed algorithm narrows from $\text{64.5}\%$ to $\text{17.5}\%$. To explain this behavior, Fig.~\ref{fig_S3_HRIS_side}(b) shows the number of active and passive elements selected by the proposed algorithm. As $M_y$ increases from $\text{40}$ to $\text{65}$, the number of activated elements drops monotonically from $\text{55}$ to only $\text{5}$. This demonstrates the trade-off between passive aperture gain and active amplification. When the array is large, massive passive elements provide sufficient beamforming gain to mitigate cascaded path loss. Therefore, the algorithm deactivates active elements to save circuit power. Furthermore, although a massive passive aperture mitigates cascaded path loss, the mode-dependent HRIS circuit power consumption still increases due to the static circuit power of the newly added passive elements, highlighting the tradeoff between passive aperture gain and circuit power consumption.

\begin{figure}[!t]
\centering
\includegraphics[width=3.7in,trim=0 10 0 25, clip]{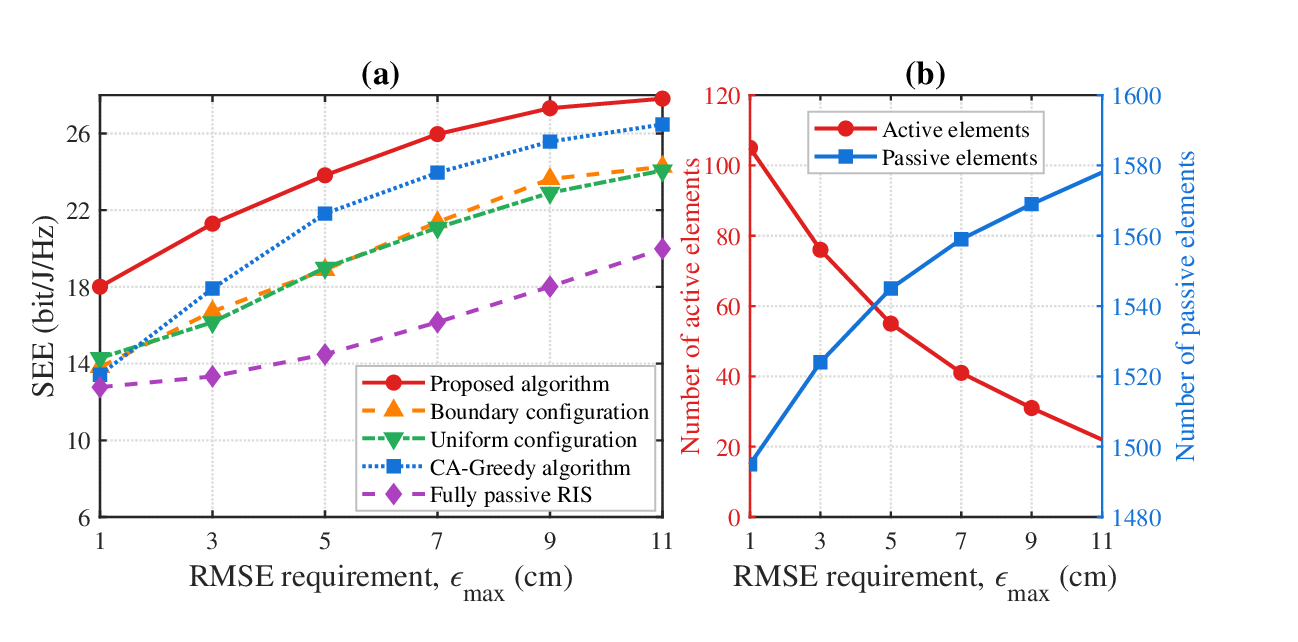}
\caption{SEE performance versus the localization RMSE requirement $\epsilon_{\max}$: (a) SEE; (b) optimized numbers of active and passive elements.}
\label{fig_S4_RMSE}
\vspace{-2mm}
\end{figure}

Fig.~\ref{fig_S4_RMSE} investigates the system SEE under
various CRB-based localization RMSE requirements
$\epsilon_{\max}\in[\text{1},\text{11}]$~cm.
In Fig.~\ref{fig_S4_RMSE}(a), the SEE of all schemes increases
with $\epsilon_{\max}$, and the proposed algorithm achieves
the highest SEE.
Its SEE increases more slowly at larger thresholds, indicating
smaller improvements from further relaxing the localization
requirement.
Fig.~\ref{fig_S4_RMSE}(b) shows that the optimized number of
active elements decreases as $\epsilon_{\max}$ increases,
while the number of passive elements increases within the
fixed HRIS aperture.
Under stricter localization requirements, the proposed
algorithm uses more active elements to support target
illumination and echo reception.
Relaxing the localization requirement allows more elements
to operate in passive mode, reducing the mode-dependent
circuit power consumption and helping to improve SEE.
These results show the benefits of adapting the active-element
configuration to balance localization accuracy, secure
transmission, and power consumption.

%\textwidth
\begin{figure}[!t]
  \centering
  \begin{minipage}[t]{0.48\linewidth}
   % \vspace{0pt}
    \centering
    \includegraphics[width=1.9in, trim=15 5 0 30, clip]{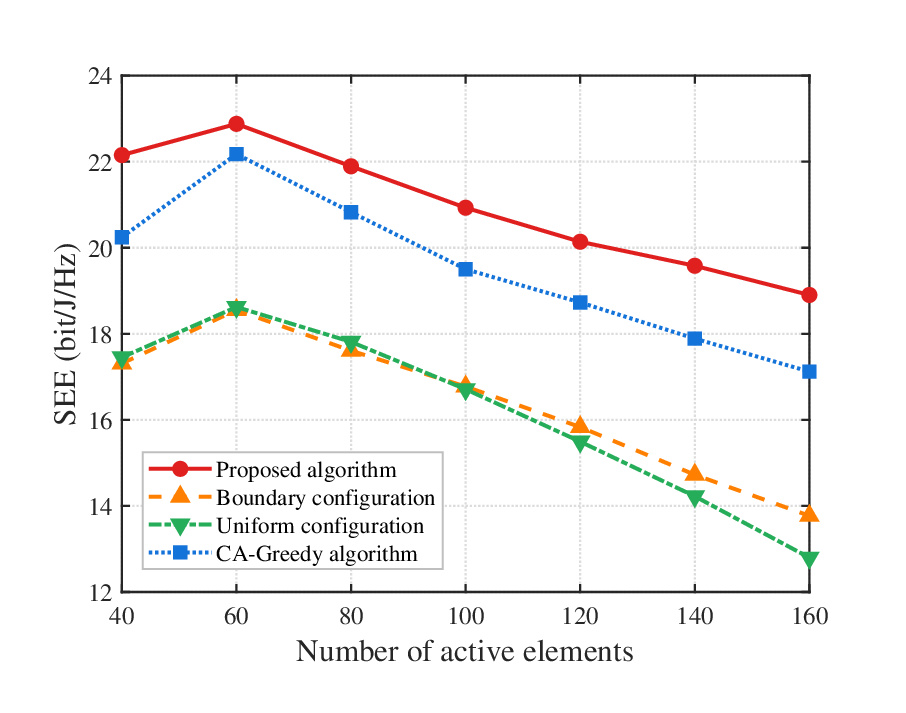} % 替换为Fig. 6文件名
    \caption{SEE performance versus a fixed number of active elements.}
    \label{fig_S5_fixed_number}
    \vspace{-3mm}
  \end{minipage}
  \hfill
  \begin{minipage}[t]{0.48\linewidth}
    %\vspace{0pt}
    \centering
    \includegraphics[width=1.9in,trim=15 5 0 30, clip]{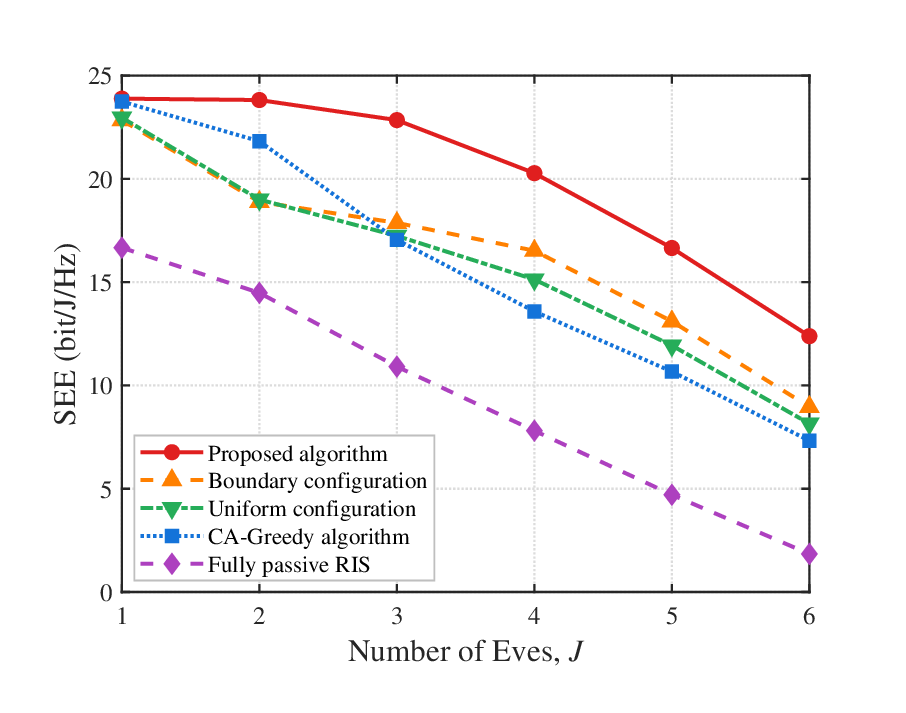}
    \caption{SEE  performance versus number of Eves $J$.}
    \label{fig_S6_J}
  \end{minipage}
  \vspace{-3mm}
\end{figure}

%\begin{figure}[!t]
%\centering
%\includegraphics[width=3.2in,trim=0 0 0 0, clip]{Figures/fig_S5_Pa.eps}
%\caption{System SEE versus circuit power consumption per active element $P_{\mathrm{a}}$.}
%\label{fig_S5_Pa}
%\end{figure}

Fig.~\ref{fig_S5_fixed_number} investigates the system SEE
with fixed numbers of active elements ranging from
$\text{40}$ to $\text{160}$.
The SEE of all schemes first increases and then decreases,
reaching its highest value at $\text{60}$ active elements
among the tested counts.
At larger counts, the benefit of additional amplification
does not compensate for the increased circuit power consumption.
For a given active-element count, all schemes have the same
mode-dependent circuit power consumption.
The proposed algorithm achieves the highest SEE at each fixed number of active elements, showing that the spatial distribution of active elements also plays an important role under near-field propagation, since elements at different locations experience distinct phase and attenuation responses and thus contribute differently to secure transmission and sensing.

%\begin{figure}[!t]
%\centering
%\includegraphics[width=3.2in,trim=0 0 0 0, clip]{Figures/fig_S6_J.eps}
%\caption{SEE versus number of Eves $J$.}
%\label{fig_S6_J}
%\end{figure}

Fig.~\ref{fig_S6_J} investigates the system SEE under various numbers of Eves $J \in [\text{1}, \text{6}]$ with $K = \text{2}$. The SEE of all schemes decreases monotonically as $J$ increases, because additional Eves introduce more eavesdropping channels and impose more localization constraints. Across the evaluated range, the proposed algorithm consistently achieves the highest SEE. When $J$ increases from $\text{1}$ to $\text{2}$, the SEE of the proposed algorithm decreases marginally, showing a small loss for this additional Eve. However, as $J$ continues to grow, heuristic methods like CA-Greedy experience rapid performance drops, restricting their capability to adapt to the coupled secrecy and localization constraints imposed by multiple spatially distributed targets.

\vspace{-2mm}
\section{Conclusion}
This paper developed a joint active-element configuration and transmission design framework to maximize the SEE of a large-aperture HRIS-assisted secure near-field ISAC system. By adapting both the number and spatial distribution of active elements to element-dependent near-field channels, the proposed framework balances bidirectional amplification gains and power consumption while supporting secure transmission and 3D target localization. Numerical results demonstrated higher SEE with lower BS transmit power than the considered fixed-layout HRIS schemes and the fully passive RIS baseline. The results also revealed that adapting active amplification to aperture size and localization requirements effectively balances performance gains and power consumption. These findings highlight the potential of adaptive HRIS architectures for energy-efficient and secure 6G applications, such as high-frequency indoor hotspots and smart industrial environments. Future work may consider mobile target tracking, robust designs under imperfect channel state information, and mixed near- and far-field propagation scenarios.

\vspace{-2.5mm}
\bibliographystyle{IEEEtran}
\bibliography{IEEEabrv,reference}

\vfill

\end{document}